\documentclass[aip,pop,floatfix,showpacs,amsmath,amssymb]{revtex4-1} 
\usepackage{graphicx}
\usepackage{bm}
\newcommand{\dif}{\mathrm{d}}

\newcommand{\mi}{\mathrm{i}}

\begin{document}
\title{Convective radial energy flux due to resonant magnetic perturbations and magnetic curvature at
the tokamak plasma edge} 

\author{F.A. Marcus} 
\email[]{albertus@if.usp.br}
\affiliation{Institute of Physics at University of S\~ao Paulo C.P. 66318, 05315-970 S\~ao Paulo, S.P., Brazil}
\affiliation{Aix-Marseille Universit\'e, CNRS, PIIM UMR 7345, 13397 Marseille Cedex 20, France}
\author{P. Beyer}
\author{G. Fuhr}
\author{A. Monnier}
\author{S. Benkadda}
\affiliation{Aix-Marseille Universit\'e, CNRS, PIIM UMR 7345, 13397 Marseille Cedex 20, France}


\begin{abstract}
With the resonant magnetic perturbations (RMPs) consolidating as an important tool to control the transport barrier relaxation, the mechanism on how they work is still a subject to be clearly understood. In this work we investigate the equilibrium states in the presence of RMPs for a reduced MHD model using 3D electromagnetic fluid numerical code (EMEDGE3D) with a single harmonic RMP (single magnetic island chain) and multiple harmonics RMPs in cylindrical and toroidal geometry. Two different equilibrium states were found in the presence of the RMPs with different characteristics for each of the geometries used. For the cylindrical geometry in the presence of a single RMP, the equilibrium state is characterized by a strong convective radial thermal flux and the generation of a mean poloidal velocity shear. In contrast, for toroidal geometry the thermal flux is dominated by the magnetic flutter. For multiple RMPs, the high amplitude of the convective flux and poloidal rotation are basically the same in cylindrical geometry, but in toroidal geometry the convective thermal flux and the poloidal rotation appear only with the islands overlapping of the linear coupling between neighbouring poloidal wavenumbers $m$, $m-1$, $m+1$. 

\end{abstract}


\maketitle 

\section{Introduction}

The key to improve plasma confinement in tokamaks at high temperature and density lies on the transport barriers achieved in so called H-mode regimes, first discovered on ASDEX \cite{Wagner_PRL82} in the early 1980's. This high energy confinement mode (H-mode) with a steep pressure gradient at the edge represents out of few percent of the plasma radius in toroidal magnetic systems. In this regime the plasma pressure drops sharply over a narrow layer in the middle of the plasma edge. Once the H-mode is settled, this pressure gradient tends to increase with time until the peeling-ballooning instability leads to a turbulent state triggering the Edge Localized Modes (ELMs), which release the transport barriers in a time shot compared to the the energy confinement time $\tau_C$. \cite{Connor_PoP98,Groeb_NF09}. ELMs are periodic fast bursts of hot dense plasma on a fast time scale ($\sim 25-300 \mu$s) and low amplitude, followed by expulsion of edge plasma and MHD activity, leading to confinement degradation \cite{Connor_PPCF98}. Along with the fast barrier relaxation event, a turbulent transport through the barrier strongly increases and the pressure gradient drops down reaching a new configuration and then the barrier builds up again on a slow collisional time scale. 
  
Turbulence simulations of transport barrier relaxations at the tokamak plasma edge have revealed that the control of such relaxations by RMPs is attributed to a local erosion of the barrier \cite{LecB_PRL09,LecB_NF10,BeyS_PPCF11}. This erosion at the resonance position is known to be caused - at least partly - by the enhancement of the radial heat flux in presence of the RMP due to the strong parallel heat flow along perturbed field lines as is the so called magnetic flutter flux \cite{Fitz_PoP95}. However, in the presence of magnetic curvature, an additional transport mechanism exists and is linked to stationary convection cells associated with the magnetic islands induced by RMPs\cite{LecB_PRL09,LecB_NF10}. In presence of a mean poloidal velocity shear, this additional convective transport can be considerably high and even larger than the thermal flux from the magnetic flutter flux\cite{BeyS_PPCF11}. This previous result has been obtained in the frame of an electrostatic model.

We show here by numerical simulations in the frame of a 3D electromagnetic fluid
turbulence model (EMEDGE3D)\cite{FuhB_PRL08} and in the basic situation without turbulent fluctuations and without imposed mean velocity shear that two different equilibrium plasma states exists in presence of a RMP. The first regime is characterized by the absence of mean poloidal flow and a low level of convective transport. The second regime shows mean poloidal rotation and large convective transport. Here we show how these two equilibria depend on magnetic curvature and RMP amplitude.
Interestingly, in the case of cylindrical curvature, the simple equilibrium
without mean poloidal rotation and without considerable convective transport is
found to be unstable such that the plasma evolves self-consistently to the
poloidally rotating state where large convective transport is present. In the
case of toroidal curvature, the simple equilibrium is found to be stable. A
detailed stability analysis is presented.

In section\ \ref{sec:3Dmodel}, the dimensionless 3D model of partial differential equations is presented with the profiles used for the code EMEDGE3D. In section\ \ref{sec:equilibrium}, the helical equilibrium states of the plasma driven by the RMP coils is derived for the cylindrical case and compared with simulations, for then analyse the results obtained in terms of the islands width in section\ \ref{sec:rotation_transition}. We show the different equilibrium states obtained with the toroidal geometry with multiple harmonics RMPs in section\ \ref{sec:toroidal}, then  we detach the main results of the work in section\ \ref{sec:conclusion} as a conclusion.   

\section{3D Plasma-RMP Model}
\label{sec:3Dmodel}

Following Beyer\cite{BeyB_PPCF07} and Fuhr\cite{FuhB_PRL08}, we introduce three reduced normalized magneto-hydrodynamical PDEs with source $J_{RMP}$ for the strength of the RMP currents. The equations are used to study the spatio-temporal behaviour of the three-dimensional fields of plasma pressure $p$, electrostatic potential $\phi$ and magnetic flux $\psi$.
\begin{eqnarray}
\partial_t \nabla_\perp^2 \phi + \left\{ \phi, \nabla_\perp^2 \phi \right\} = -
\alpha^{-1} {\nabla_\parallel} \nabla_\perp^2 \psi - {\bf G} p + \nu
\nabla_\perp^4 \phi \;, \label{eq:model_phi}\\
\partial_t p + \left\{ \phi, p \right\} = \delta_c {\bf G} \phi + \chi_\parallel
{\nabla_\parallel}^2 p + \nabla_\perp \cdot \left[ \chi_\perp(x) \nabla_\perp p
\right] + S(x) \;, \label{eq:model_p}\\
\partial_t \psi = - {\nabla_\parallel} \phi + \alpha^{-1} \nabla_\perp^2 \psi -
\alpha^{-1} J_\mathrm{RMP} \label{eq:model_psi}\;.
\end{eqnarray}
In toroidal coordinates $(r, \theta, \varphi)$ and in a slab geometry $(x, y,
z)$ in the vicinity of a reference surface $r = r_0$ at the plasma edge, i.e.\
$x = \left( r - r_0 \right)/ \xi_\mathrm{bal}$, $y = r_0 \theta /
\xi_\mathrm{bal}$, $z = R_0 \varphi / L_s$, the normalized operators are
\begin{eqnarray*}
\nabla_\parallel & = & \partial_z + \left( \kappa / q_0 - x \right) \partial_y -
\left\{ \psi, \ \cdot \ \right\} \quad \mbox{with} \quad \kappa = \frac{L_s
r_0}{R_0 \xi_{\mathrm{bal}}} \;, \\
\nabla_\perp^2 & = & \partial_x^2 + \partial_y^2
\;, \\
\left\{ \phi, \ \cdot \ \right\} & = & \partial_x \phi \partial_y -
\partial_y \phi \partial_x \;, \\ 
{\bf G} & = & \left\{ \begin{array}{ll} \sin \theta \, \partial_x + \cos
\theta \, \partial_y \;, & \mbox{in case of toroidal curvature} \\
g_0 \partial_y \;, & \mbox{in case of cylindrical curvature} \end{array} \right. \;.
\end{eqnarray*}
Here, $q_0 = q(r_0)$ is the safety factor at the reference surface, $R_0$ is the
major radius of the magnetic axis and $L_s$ is the magnetic shear length. Lengths
parallel ($\parallel$) and perpendicular ($\perp$) to the unperturbed magnetic
field are normalized by $L_s$ and $\xi_\mathrm{bal}$, respectively, and time is
normalized by $\tau_\mathrm{int}$, where the resistive ballooning radial correlation length $\xi_\mathrm{bal}$ and the interchange time $\tau_\mathrm{int}$ are given by
\begin{displaymath}
\xi_\mathrm{bal} = \left( \frac{L_p}{\tau_e c_s} \frac{m_e}{m_i}
\right)^\frac{1}{2} \frac{L_s}{L_p} \left( \frac{L_p}{R_0} \right)^\frac{1}{4}
\rho_s \ \mbox{and} \quad \tau_\mathrm{int} =  \frac{\left(L_p R_0\right)^\frac{1}{2}}{\sqrt{2} c_s} \;, 
\end{displaymath}
where $m_e / m_i$ is the ratio of the electron to the ion mass, and $\tau_e$,
$c_s$, $L_p$, $\rho_s$ are reference values of the electron collision time,
the sound speed, the pressure gradient length, and the ion Larmor radius at
electron temperature, respectively. For a collisional tokamak plasma edge, one
typically finds $\xi_\mathrm{bal} \sim \rho_s$ and $\tau_\mathrm{int} \sim 10
L_p / c_s$. 

The dimensionless ballooning pressure gradient coefficient $\alpha = (\beta / L_p) (L_s^2 / R_0)$ is similar to the normalized pressure gradient $\alpha_{\text{MHD}}$ widely used in MHD tokamak stability theory [$\alpha_{\text{MHD}} = \alpha (q_0 R_0 / L_s)^2$], where $\beta$ is the ratio of plasma pressure to magnetic pressure. The parallel and perpendicular heat conductivity coefficients $\chi_\parallel$ and $\chi_\perp$ are normalized by $L_s^2 / \tau_\mathrm{int}$ and $\xi_\mathrm{bal}^2 / \tau_\mathrm{int}$, respectively. Therefore, a ratio of the normalized coefficients of $\chi_\parallel / \chi_\perp \sim 1$, corresponds to a ratio of the dimensional coefficients of $L_s^2 / \xi_\mathrm{bal}^2 \sim 10^7-10^8$. In the present simulations, we use $\alpha = 0.1$, $\nu = \chi_\perp = 0.93$, $\chi_\parallel = 1$ and the curvature parameter $\delta_c = \frac{5}{3}\, 2 L_p / R_0$ is set to $\delta_c = 0.01$.

\begin{figure}[h]
\centerline{\includegraphics[scale=0.7]{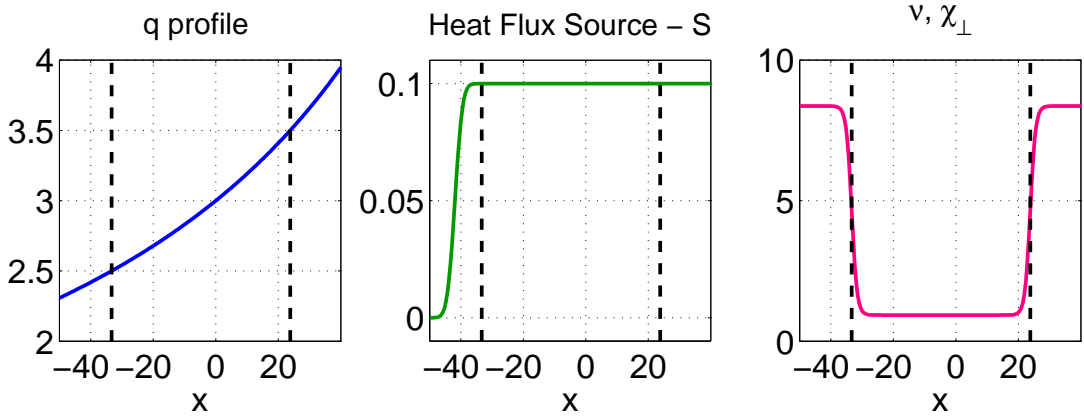}}
\caption{Radial profiles of the safety factor (a), the incoming energy flux (b) and perpendicular diffusivity and viscosity (c) as a function of the normalized radial coordinate $x$. The vertical dashed lines delimit the main computational domain, which is between $x=-33.3$ and $x=23.8$ and where $q=2.5$ and $q=3.5$, respectively. The left part is associated to the incoming energy flux from the core and the right to the external magnetic perturbations, both defined as buffer zone. 
\label{fig:radial_profiles}}
\end{figure}

The main computational domain corresponds to the volume delimited by the toroidal surfaces characterized by $q = 2.5$ and $q = 3.5$, respectively, and including the reference surface $q = q_0 = 3$. Here, a linear $1/q$ profile is assumed, and $\xi_\mathrm{bal} / r_0 = 1/500$, $L_s/R_0 = 1$ for the reference parameters. The complete computational domain is slightly larger and delimited by $x_\mathrm{min} < x_{q=2.5}$ and $x_\mathrm{max} > x_{q=3.5}$. The source $S$ is located in the inner buffer zone $x_\mathrm{min} < x < x_{q=2.5}$ and gives rise to a constant incoming energy flux, $Q_\mathrm{tot} =\int_{x_\mathrm{min}}^{x_{q=2.5}} S\,\dif x$ from the plasma center into the main computational domain. The helical current $J_0(x)$ 

\begin{figure}[h]
\centerline{\includegraphics[width=0.4\textwidth]{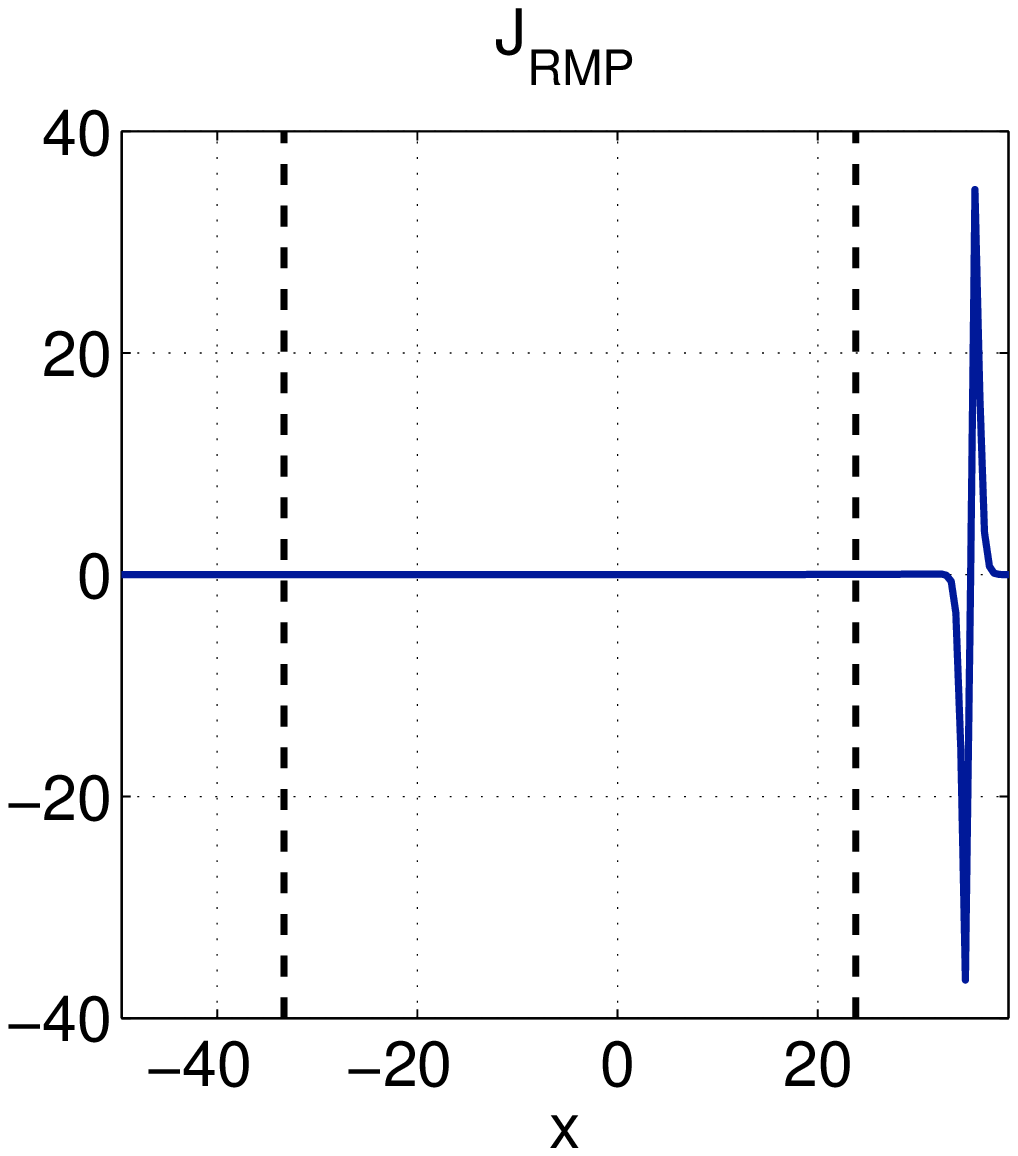}
\includegraphics[width=0.4\textwidth]{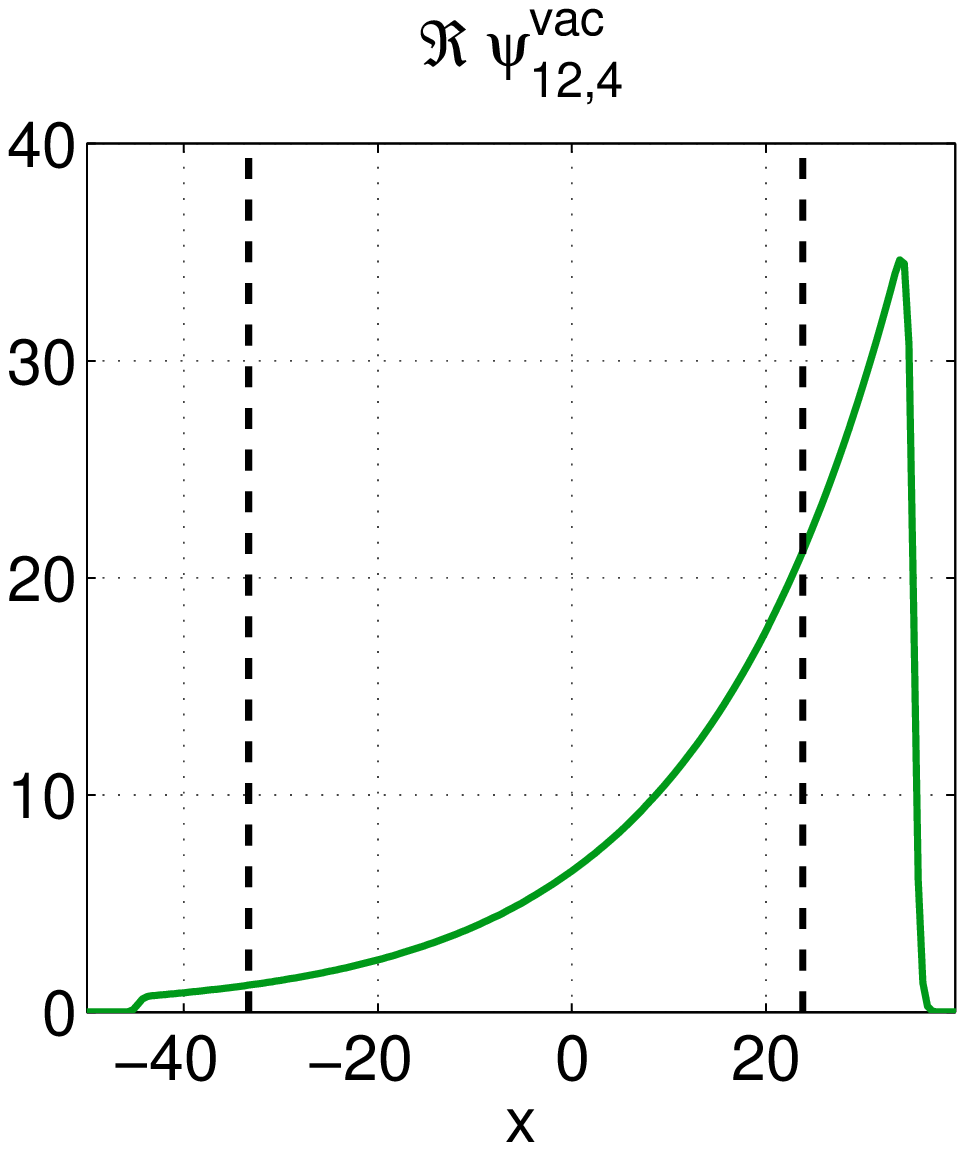}}
\caption{Radial profiles of (a) the amplitude $J_0(x)$ of the external helical current and (b) the resulting vacuum magnetic perturbation $\psi_{12,4}$. \label{fig:geometry_J_psi}}
\end{figure}
\begin{equation}
J_\mathrm{RMP} = J_0 (x) \cos \zeta \label{eq:current_source}
\end{equation}
is located in the outer buffer zone $x_{q=3.5} < x < x_\mathrm{max}$ (Fig.\ \ref{fig:geometry_J_psi}a). Here, the RMP coil structure is a function of the helical angle
\begin{displaymath}
\zeta = m_0 \theta - n_0 \varphi = \frac{m_0 \xi_{\text{bal}}}{r_0} y - \frac{n_0 L_s}{R_0} z .
\end{displaymath}
Here we choose the regime RMP coil with $(m_0, n_0) = (12,4)$, which the external current induces a magnetic perturbation resonant at $q = q_0 = 3$ as shown in Fig.\ \ref{fig:geometry_J_psi}b.

The pressure profile $\bar{p}(x, t) = \left\langle p \right\rangle_{y,z}$
evolves self consistently according to the energy transport equation [the
toroidal and poloidal average $\left\langle \cdot \right\rangle_{yz}$ of
(\ref{eq:model_p})],
\begin{equation}
\partial_t \bar{p} = - \partial_x \left( Q_\mathrm{conv} + Q_\mathrm{coll} +
Q_{\delta B} \right) + S \;, \label{eq:transport}
\end{equation}
with $Q_\mathrm{conv} = \left\langle p \,\partial_y \phi \right\rangle_{y,z}$, \
$Q_\mathrm{coll} = - \chi_\perp \partial_x \bar{p}$, \ $Q_{\delta B} = -
\chi_\parallel \left\langle \partial_y \psi \nabla_\parallel p
\right\rangle_{y,z}$. In a stationary state, integrating Eq.\ \ref{eq:transport} in the radial direction leads to the energy flux balance 
\begin{equation}
Q_\mathrm{conv}(x) + Q_\mathrm{coll}(x) + Q_{\delta B}(x) = Q_\mathrm{tot} \quad
\mathrm{for} \quad x \ge x_\mathrm{q=2.5}  \label{eq:flux_balance}
\end{equation}
for a steady state.

\section{Equilibrium states in presence of RMP}
\label{sec:equilibrium}

\begin{figure} [h]
  \centerline{\includegraphics[width=.45\textwidth]{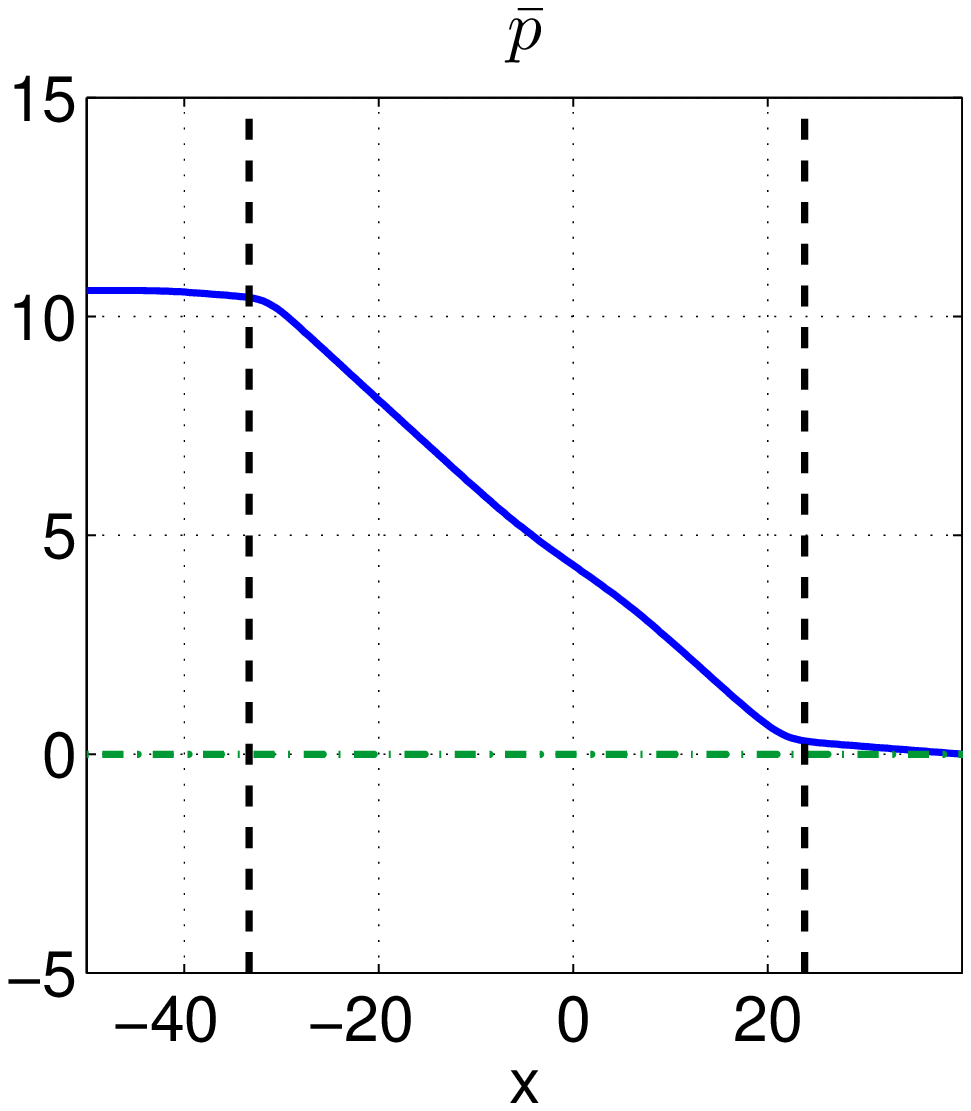}
    \hspace{-1.5cm} \includegraphics[width=.45\textwidth]{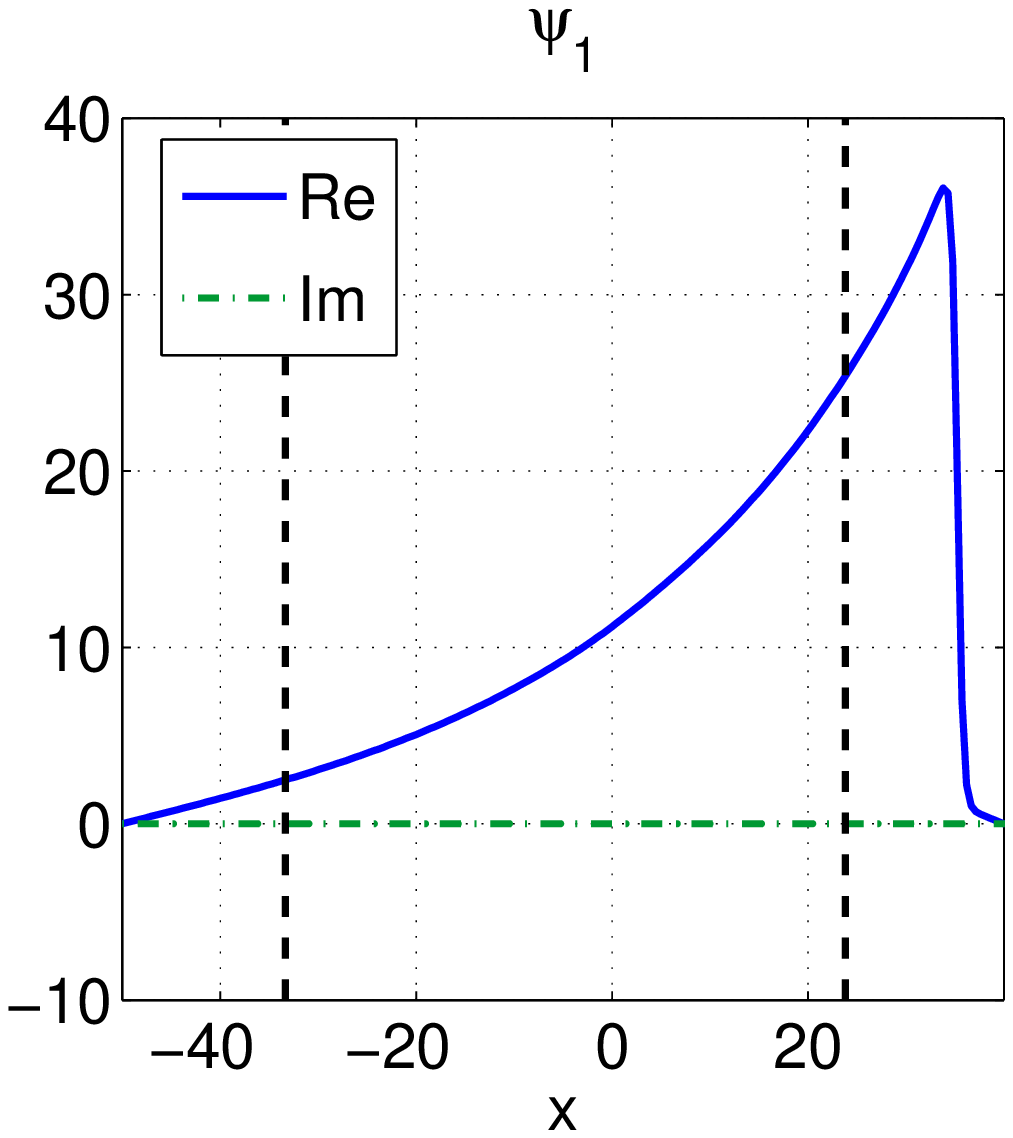}}
  \centerline{\includegraphics[width=.45\textwidth]{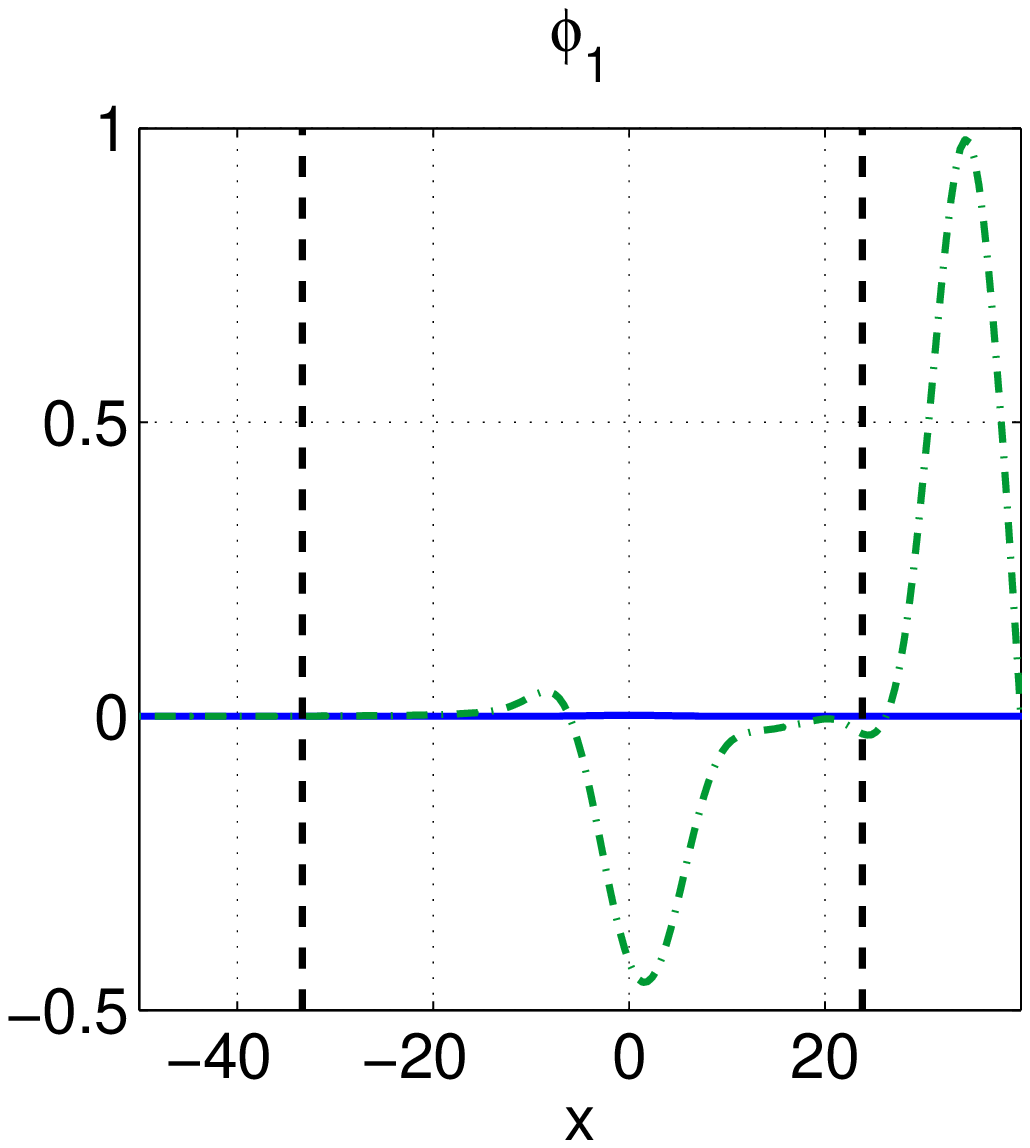}
    \hspace{-1.5cm} \includegraphics[width=.45\textwidth]{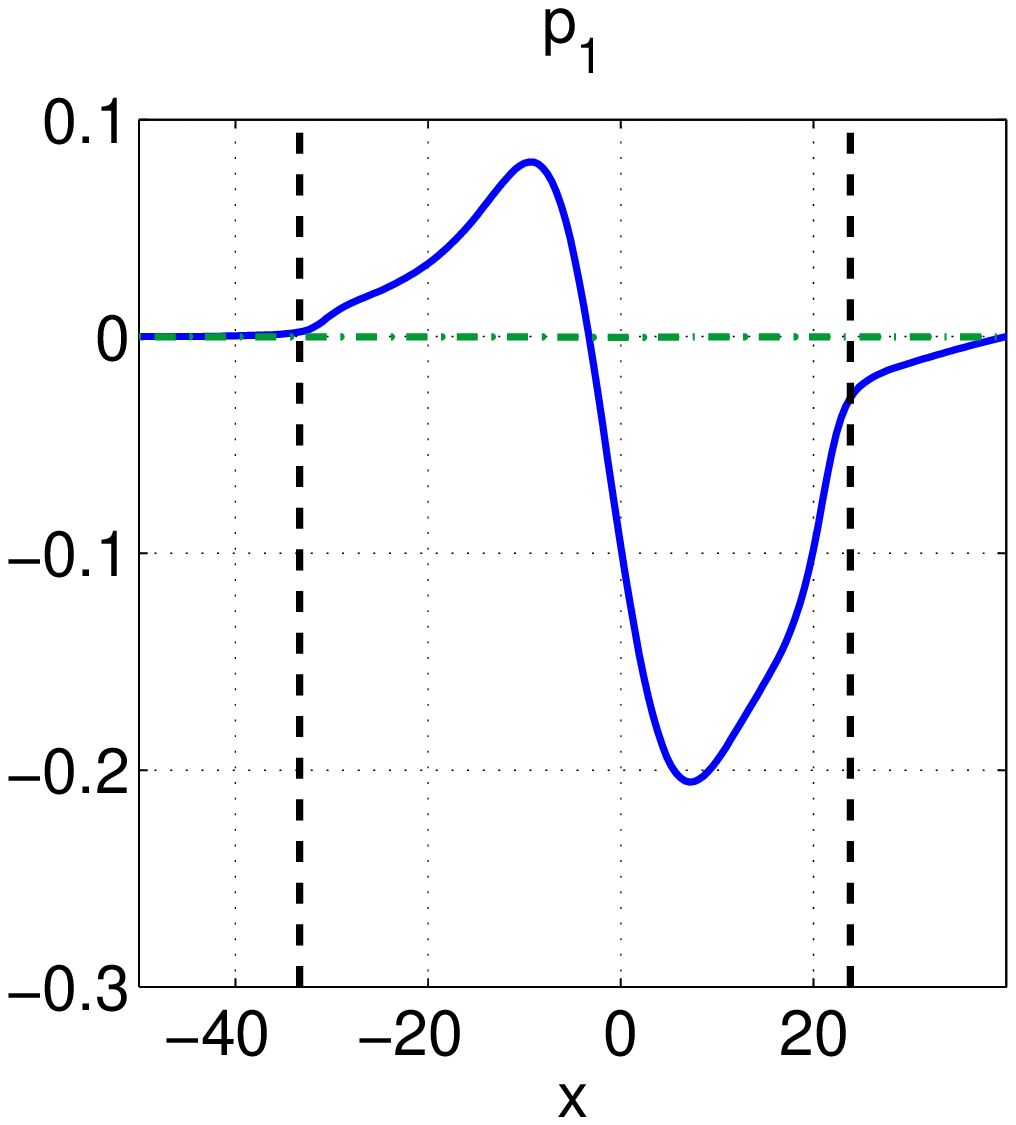}}
  \caption{Radial profiles of the fields $\bar{p}$, $\phi_1$, $p_1$ and $\psi_1$ corresponding to the calculated helical equilibrium described in Eq.\ (\ref{eq:equilibrium}). The magnetic perturbation is in phase with the external current and the plasma is not rotating ($\bar{v}_E = \partial_y \bar{\phi} = 0$), but this equilibrium is unstable according to the results obtained with numerical simulations. The dashed vertical lines mark the location of the rational magnetic surfaces with $q = 5/2$ and $q=7/2$. \label{fig:solution1}}
\end{figure}
\begin{figure}
\centerline{\includegraphics[width=0.5\textwidth]{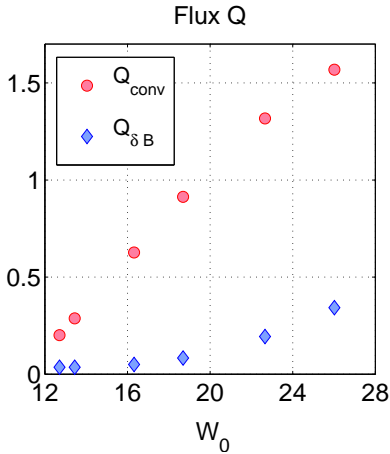}}
\caption{(Color online) Radially integrated convective and flutter fluxes for the cylindrical case, $Q_\mathrm{conv} = \int_{q=2.5}^{q=3.5} Q_\mathrm{conv}(x)\,\mathrm{d}x$ and $Q_{\delta B} = \int_{q=2.5}^{q=3.5} Q_{\delta B}(x)\,\mathrm{d}x$, as a function of a single magnetic perturbation amplitude expressed in terms of the vacuum island width $W_0$. Even for low values of external resonant magnetic perturbation $\psi_0$ (small island width $W_0$), $Q_{conv}$ is higher than $Q_{\delta B}$. \label{fig:compare_fluxes}}
\end{figure}

In the following, for a given RMP coil, two different helical equilibrium states of the plasma are obtained by the numerical calculations we first deduce analytically their main properties in the case of cylindrical curvature. In this case, there is no
linear coupling between neighbouring poloidal wavenumbers $m$, $m-1$, $m+1$, etc.,
and the equilibrium can be assumed to be of the form
\begin{equation}
\left( \begin{array}{c} \phi_\mathrm{eq} \\ p_\mathrm{eq} \\ \psi_\mathrm{eq}
\end{array} \right) = \left( \begin{array}{c} \bar{\phi}(x) \\ \bar{p}(x) \\ 0
\end{array} \right) + \left( \begin{array}{c} \phi_1 (x) \\ p_1 (x) \\ \psi_1
(x) \end{array} \right) \exp \left( \mi \zeta \right) + c.c. \;,
\label{eq:harmonics} \\
\end{equation}
where the bar and the index "1" designate respectively the axisymmetric and the $(m, n) = (m_0, n_0) = (12, 4)$ components, in which 
\begin{displaymath}
\left( \phi_1, p_1, \psi_1 \right) \equiv \left( \phi_{m_0, n_0}, p_{m_0, n_0}, \psi_{m_0, n_0} \right) \equiv \left( \phi_{12, 4}, p_{12, 4}, \psi_{12, 4} \right) \;.
\end{displaymath}
Inserting the expression (\ref{eq:harmonics}) in Eq.\
(\ref{eq:model_phi})--(\ref{eq:model_psi}) yields the following set of five coupled equations for $\bar{\phi}$, $\bar{p}$, $\phi_1$, $p_1$ and $\psi_1$,
\begin{eqnarray}
- \frac{1}{2} k_y \partial_x \Im \left[ \phi_1 \left( \partial_x^2 \!-\! k_y^2
\right) \phi_1^* \right] & = & - \frac{1}{2 \alpha} k_y \partial_x \Im \left[ \psi_1
\left( \partial_x^2 \!-\! k_y^2 \right) \psi_1^* \right] + \nu \partial_x^4
\bar{\phi} \;, \label{eq:response_phi_bar}\\
2 k_y \partial_x \Im \left( \phi_1 p_1^* \right) & = & - 2 \chi_\parallel k_y^2
\partial_x \left[ x \Re \left( \psi_1 p_1^* \right) - \left| \psi_1 \right|^2
\partial_x \bar{p} \right] + \chi_\perp \partial_x^2 \bar{p} + S \;.
\label{eq:response_p_bar} \\
\mi \left[ \partial_x \bar{\phi} \left( \partial_x^2 \!-\! k_y^2 \right) \phi_1
- \phi_1 \partial_x^3 \bar{\phi} \right] & = & - \mi g_0 p_1 + \frac{\mi}{\alpha}
x \left( \partial_x^2 \!-\! k_y^2 \right) \psi_1 + \frac{\nu}{k_y} \left(
\partial_x^2 \!-\! k_y^2 \right)^2 \phi_1 \;, \label{eq:response_phi} \\
\frac{\mi}{k_y} \left( p_1 \partial_x \bar{\phi} - \phi_1 \partial_x \bar{p}
\right) & = & \frac{\mi}{k_y} \delta_c g_0 \phi_1 - \chi_\parallel x^2 p_1 +
\chi_\parallel x \psi_1 \partial_x \bar{p} + 2 \mi \chi_\parallel \psi_1
\partial_x^2 \left( \psi_1 \Im p_1 \right) \nonumber \\ \mbox{} & & +
\chi_\parallel \left( \psi_1 \partial_x - 2 \partial_x \psi_1 \right) \partial_x
\left( \psi_1 p_1 \right) + \frac{\chi_\perp}{k_y^2} \left( \partial_x^2 \!-\!
k_y^2 \right) p_1 \;, \label{eq:response_p} \\ 
0 & = & \mi k_y x \phi_1 + \mi k_y \psi_1 \partial_x \bar{\phi} + \frac{1}{\alpha}
\left( \partial_x^2 \!-\! k_y^2 \right) \psi_1  \;. \label{eq:response_psi}
\end{eqnarray}
Here, $\Re f = (f+f^*)/2$ and $\Im f = (f-f^*)/2 \imath$. Eqs.\ (\ref{eq:response_phi_bar})--(\ref{eq:response_psi}) admit a symmetric
solution corresponding to a helical equilibrium where the induced magnetic
perturbation and the associated pressure perturbation are in phase with the
external current (\ref{eq:current_source}), i.e.\ $\Im \psi_1 = 0$, $\Im p_1 =
0$, and the potential variation is in phase quadrature, $\Re \phi_1 = 0$. In
this equilibrium, the poloidal rotation is zero, $\bar{v}_E = \partial_y
\bar{\phi} = 0$. In summary, the solutions of Eq.\ (\ref{eq:response_phi_bar})--(\ref{eq:response_psi}) reduces to
\begin{equation}
\left( \begin{array}{c} \phi_\mathrm{eq}^{(1)} \\ p_\mathrm{eq}^{(1)} \\
\psi_\mathrm{eq}^{(1)} \end{array} \right) = \left( \begin{array}{c} 0 \\
\bar{p}(x) \\ 0 \end{array} \right) + \left( \begin{array}{c} -2 \phi_1^I (x)
\sin \zeta \\ 2 p_1^R (x) \cos \zeta \\ 2 \psi_1^R (x) \cos \zeta \end{array}
\right) \;, \label{eq:equilibrium} \\
\end{equation}
where $\phi_1^I$, $p_1^R$ and $\psi_1^R$ are real fields corresponding to the
imaginary (I) and real (R) parts of $\phi_1$, $p_1$ and $\psi_1$, respectively.
Typical radial profiles of $\bar{p}$, $\phi^I$, $p^R$ and $\psi^R$ are shown in
Fig.\ \ref{fig:solution1}. Note that there is a non-vanishing convective
flux $Q_\mathrm{conv}$ associated with the equilibrium (\ref{eq:equilibrium}),
as the pressure and potential perturbations are phase shifted by $\pi/2$. 
However, in typical situations as the one illustrated in Fig.\ \ref{fig:solution1}, this
convective flux is found to be much smaller than the magnetic flutter flux $Q_{\delta B}$.
\begin{figure} 
  \centerline{\includegraphics[width=.45\textwidth]{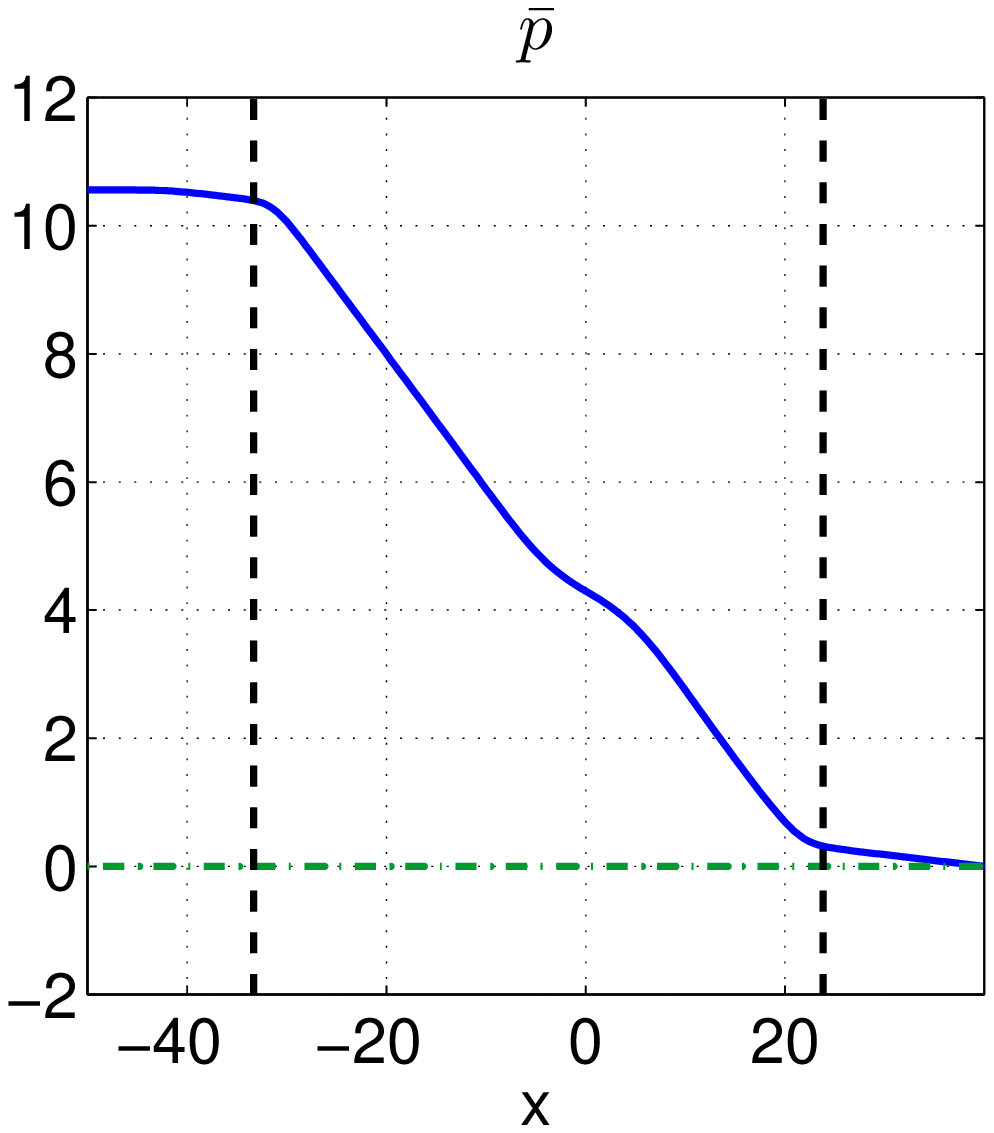}
    \hspace{-1.2cm} \includegraphics[width=.45\textwidth]{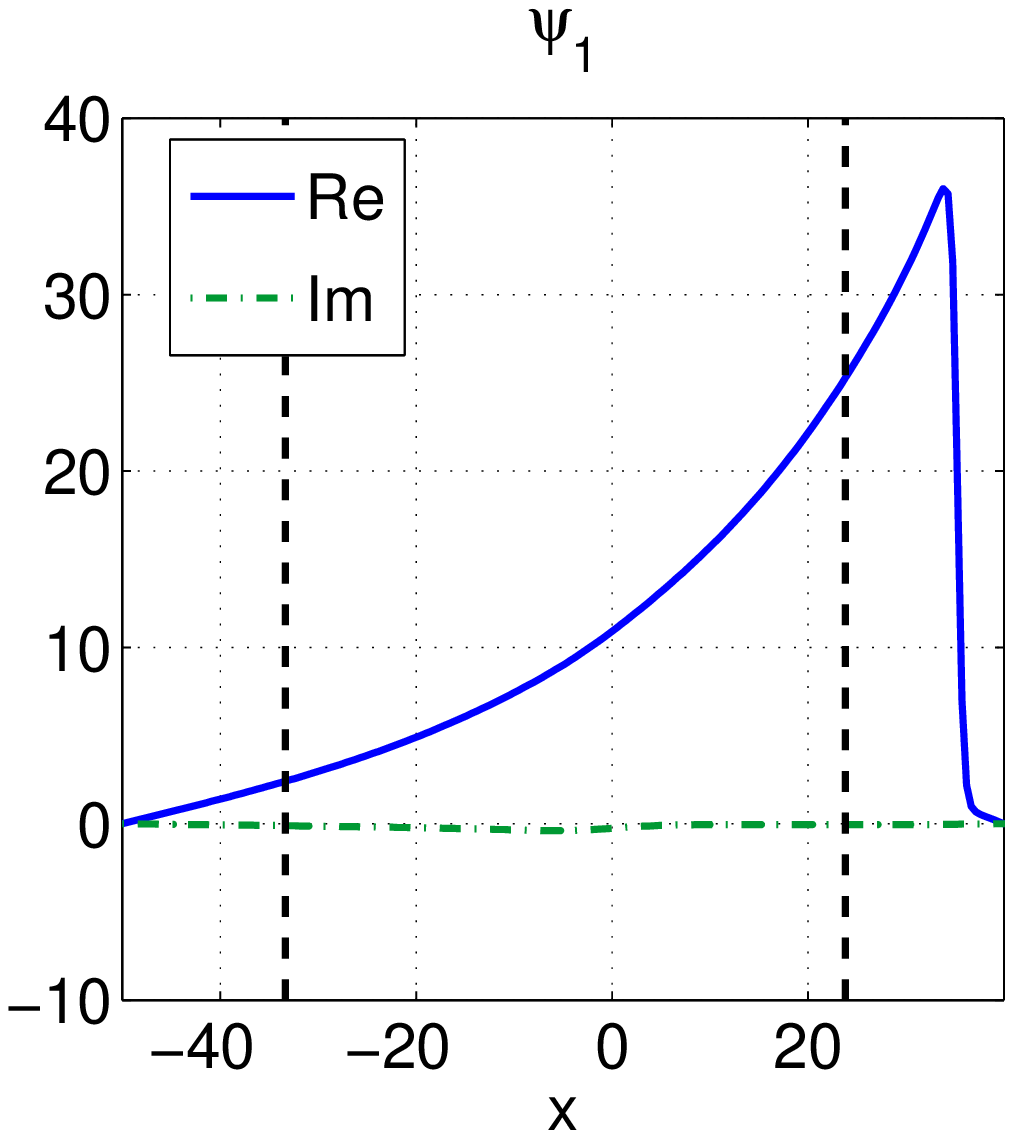}}
  \centerline{\includegraphics[width=.45\textwidth]{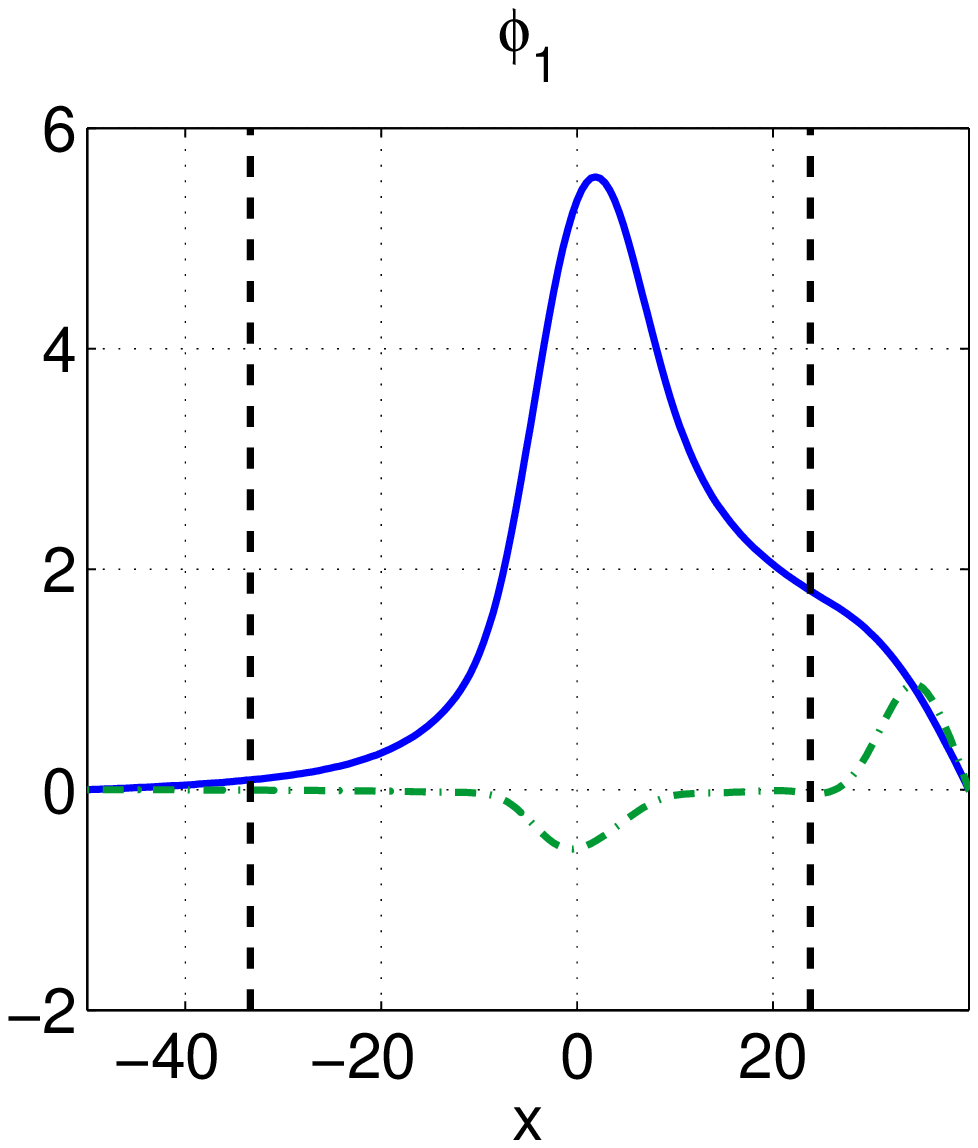}
    \hspace{-1.2cm} \includegraphics[width=.45\textwidth]{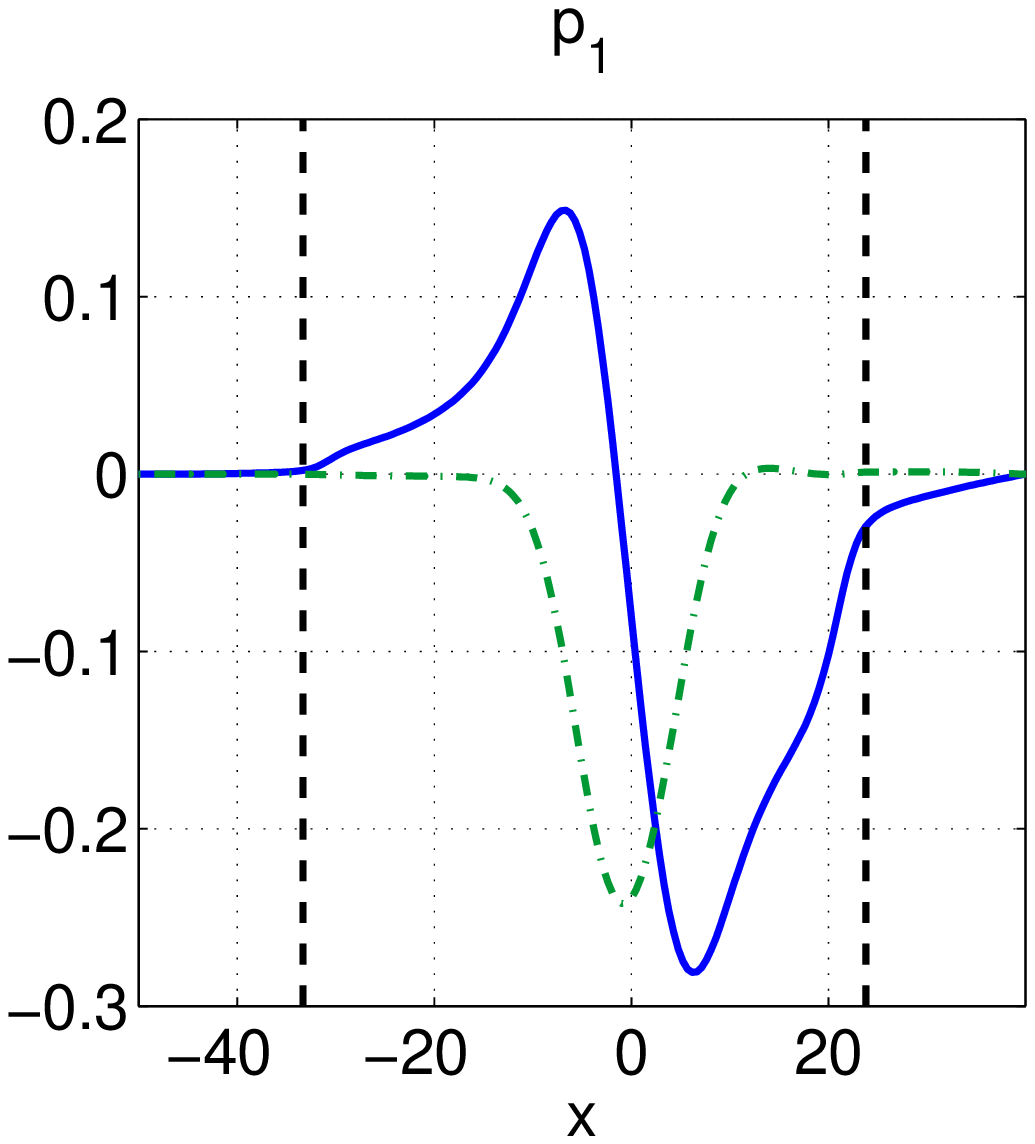}}
  \centerline{\includegraphics[width=.45\textwidth]{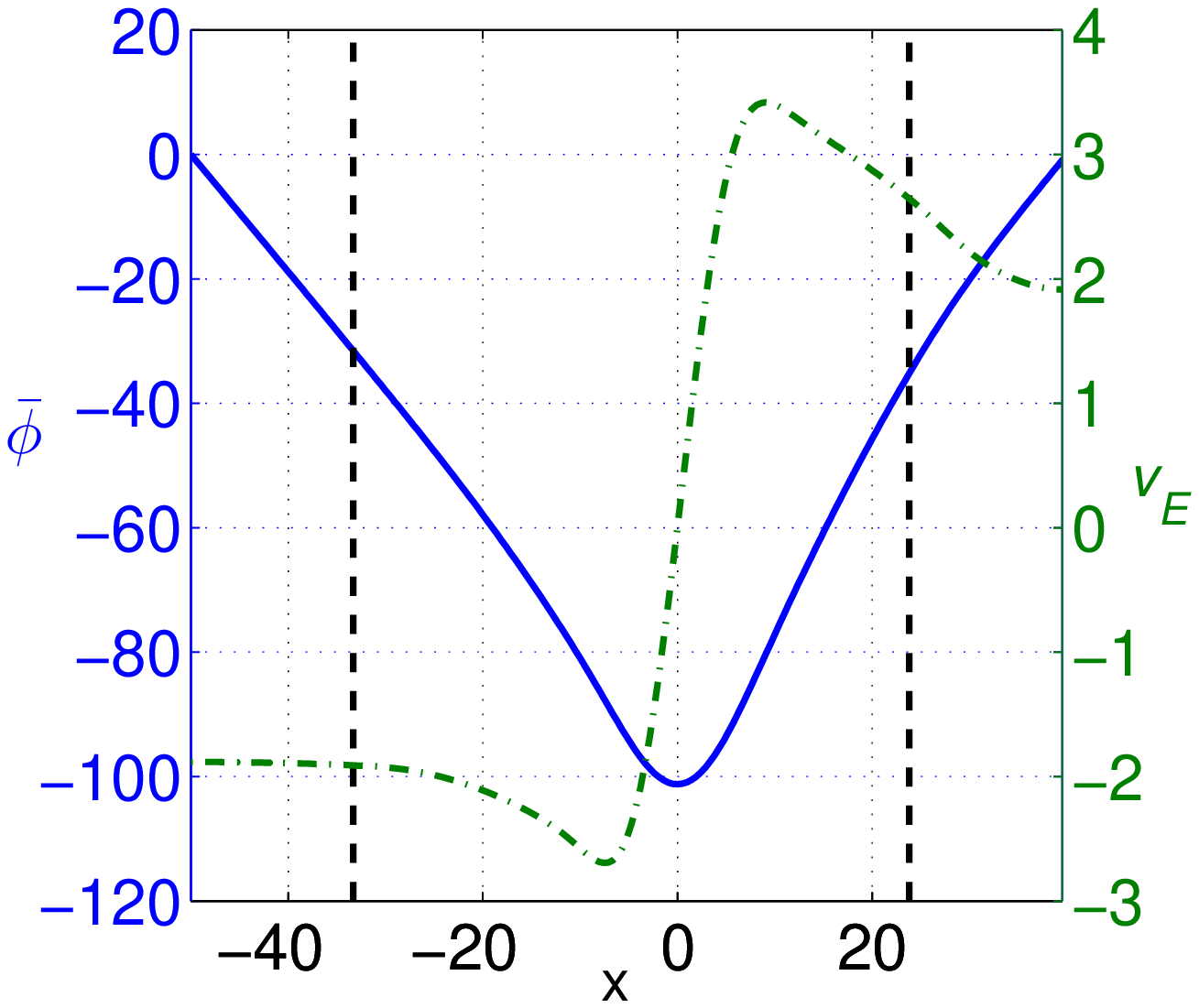}
  \hspace{-1.0cm} \includegraphics[width=.45\textwidth]{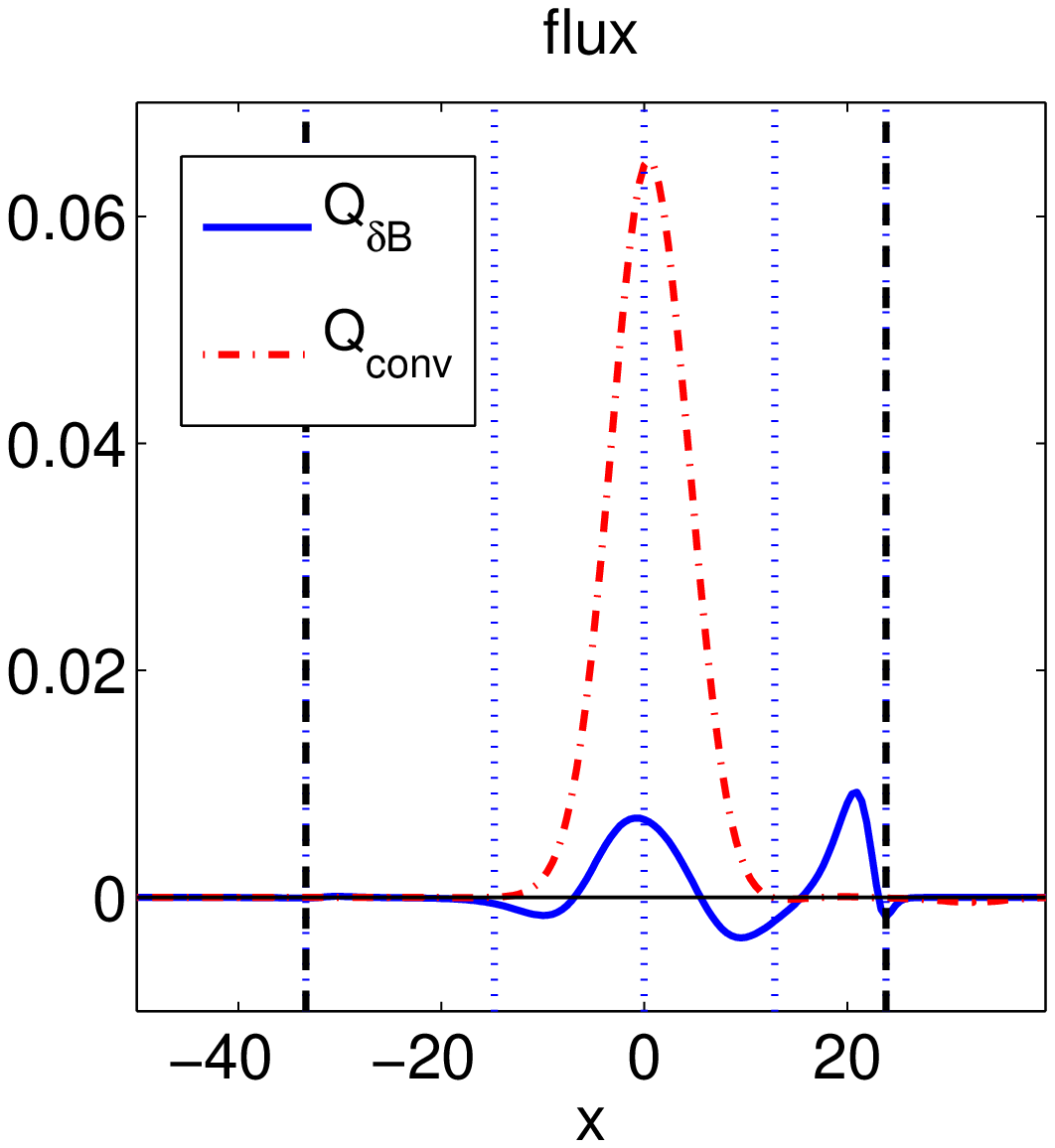}}
  \caption{(Color online) Radial profiles of the fields $\bar{p}$, $\psi_1$, $p_1$, $\phi_1$, the poloidal rotation $\bar{v}_E = \partial_y \bar{\phi}$ and the fluxes $Q_\mathrm{conv}$, $Q_{\delta B}$ corresponding to the stable helical equilibrium. The magnetic perturbation is phase shifted with respect to the external current.\label{fig:solution2}}
\end{figure}
Stable helical equilibrium states of the plasma in presence of the RMP can be
calculated numerically by integrating Eq.\ (\ref{eq:model_phi})--(\ref{eq:model_psi}) in time starting with low level noise for all fields, providing that the pressure gradient $\partial_x \bar{p}$ stays below the resistive ballooning instability limit. This is guaranteed here by choosing a sufficiently low value of the total energy flux $Q_\mathrm{tot}$.
However, with this convergence method, the symmetric equilibrium state
(\ref{eq:equilibrium}) can only be obtained when explicitly forcing no
rotation. This means that the symmetric equilibrium is unstable. This property will be
discussed in the next section.
\begin{figure}
\centerline{\includegraphics[width=0.45\textwidth]{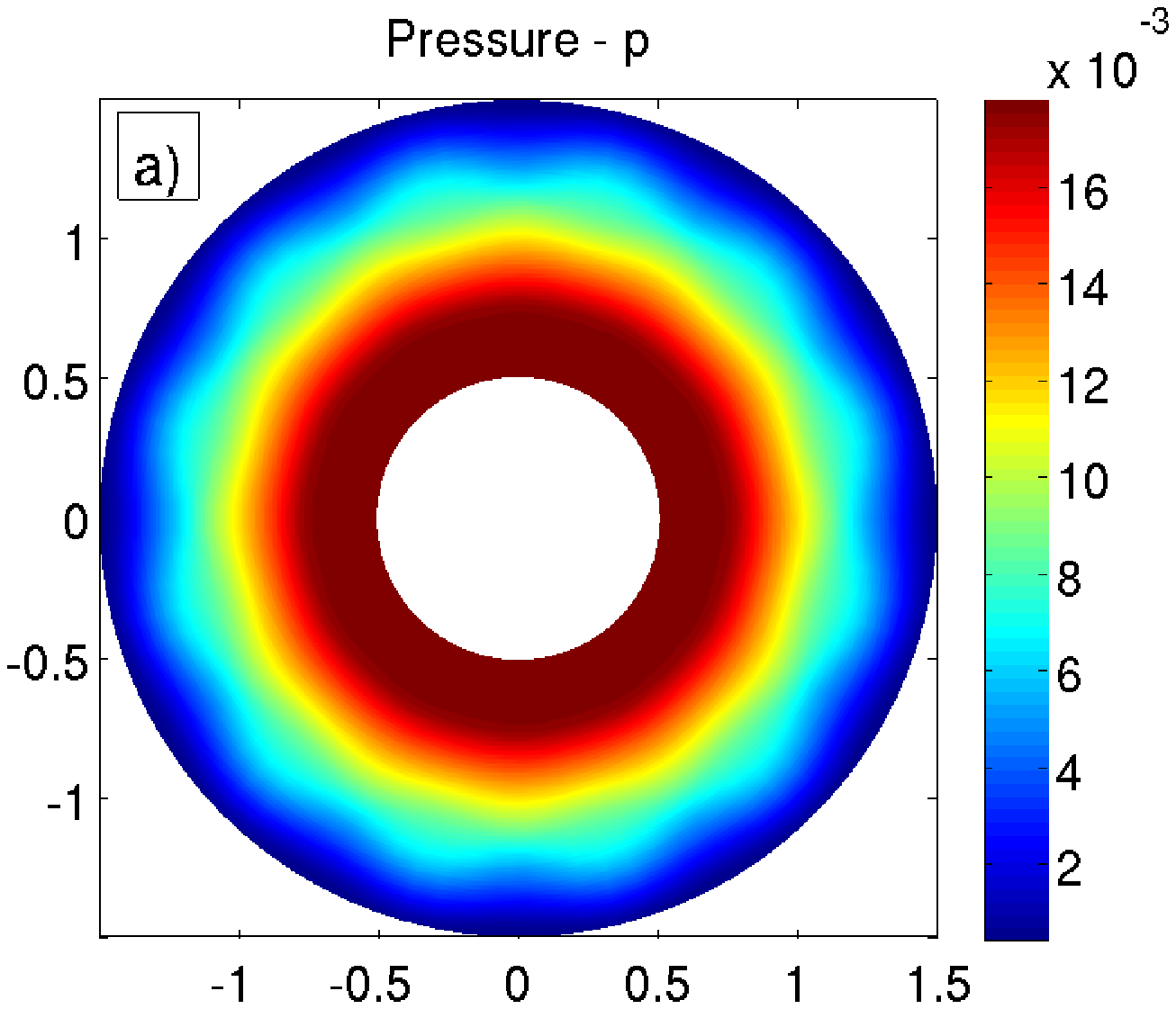}
\hfill\includegraphics[width=0.45\textwidth]{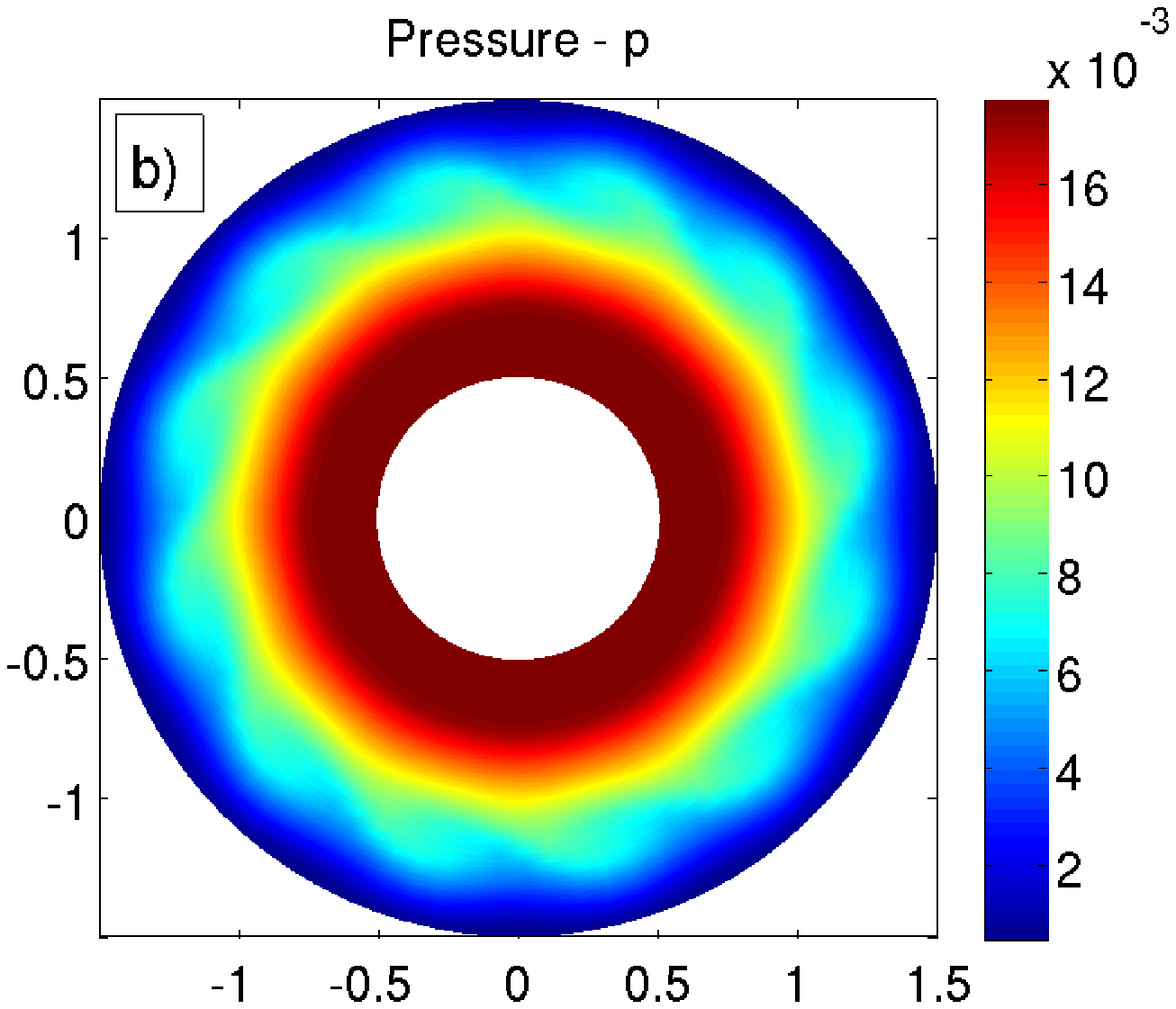}}
\centerline{\includegraphics[width=0.45\textwidth]{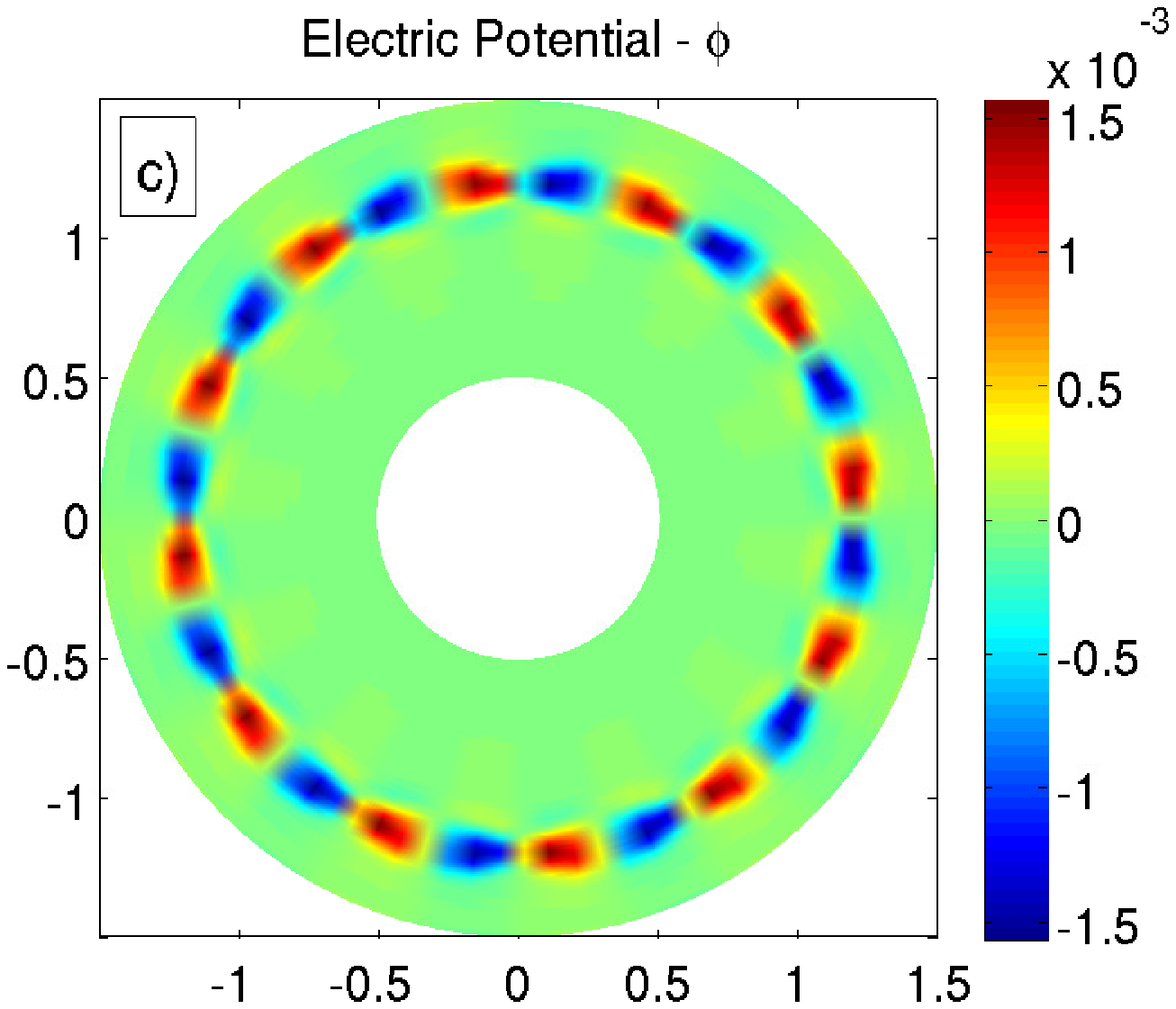}
\hfill\includegraphics[width=0.45\textwidth]{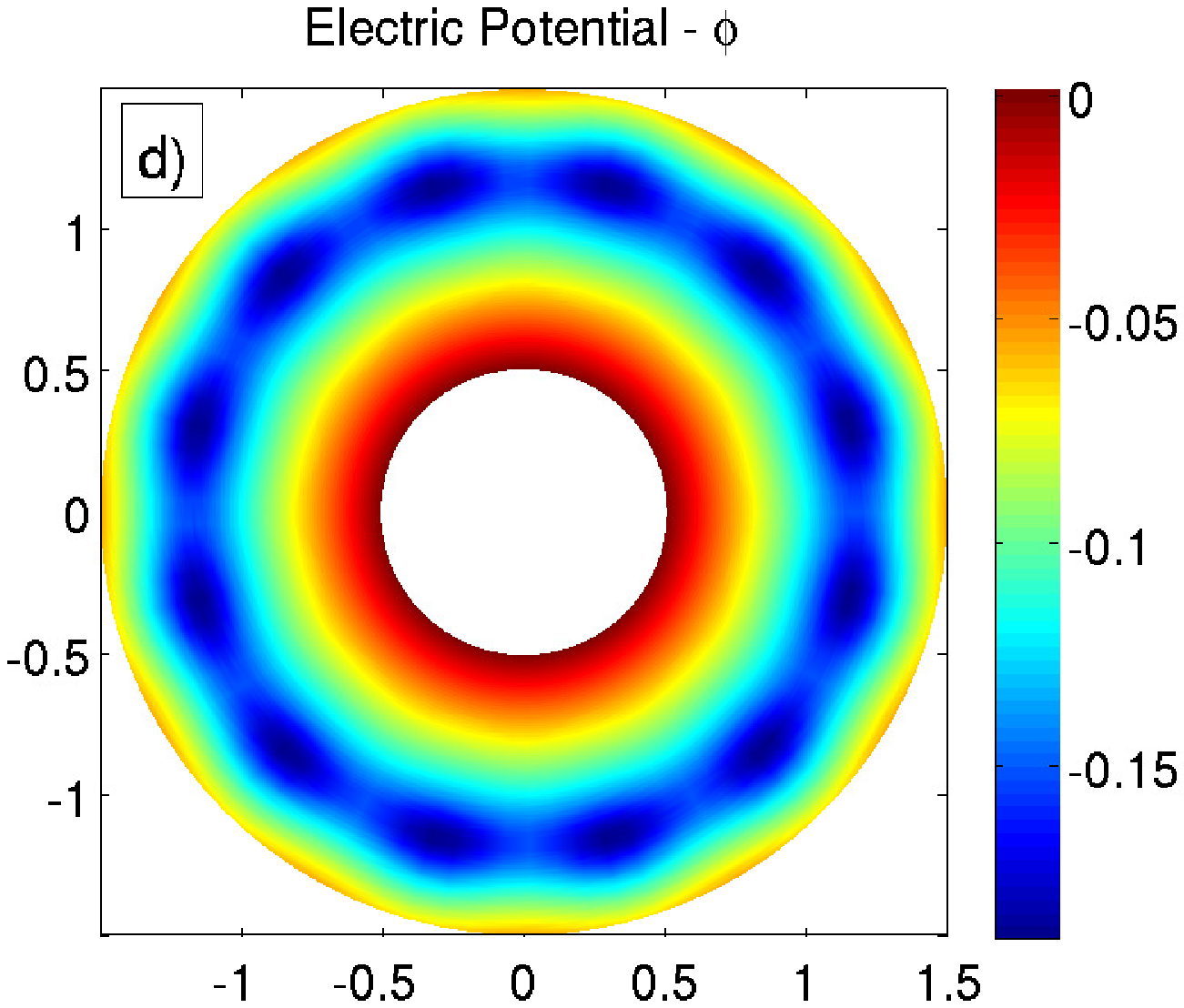}}
\caption{(Color online) Two dimensional maps of the pressure $p$ and the non-axisymmetric part of potential $\phi - \bar{\phi}$ corresponding to the two different equilibrium states in cylindrical geometry. Pictures a) and c) depict the pressure and electric potential profiles for the case described by system in equilibrium in Eq.\ (\ref{eq:equilibrium}) with the profiles shown on Fig.\ \ref{fig:solution1}. This system evolves to the state depicted on frames b) and d) with poloidal rotation, which the phase-shift between $p$ and $\partial \phi$ play the role to the high convective flux and the poloidal rotation. \label{fig:poloidal_cuts}}
\end{figure}
When the temporal evolution is calculated self-consistently including poloidal
rotation, the system evolves to a new helical stationary state
with non vanishing sheared plasma rotation. We call this new equilibrium here \textit{rotating state} and it is characterized by: A. a phase shift between the external current and the magnetic perturbation induced in the plasma and B. a large convective flux
$Q_\mathrm{conv} \gg Q_{\delta B}$. Typical radial profiles of the axisymmetric components
as well as the helical amplitudes and the fluxes are shown in Fig.\ \ref{fig:solution2}.
Note that the rotation velocity vanishes close to the resonant surface which is consistent with a nearly complete penetration of the magnetic perturbation \cite{MonF_NF14}, i.e.\ the amplitudes of the $\psi_{12,4}(x)$ at the resonance surface $x = 0$ are similar in both equilibria and close to the maximum value corresponding to the vacuum case. We may summarise the fields in the rotating steady state as
\begin{equation}
\hat{\psi}_{12,4}^{\text{rotation state}} \approx \hat{\psi}_{12,4}^{\text{symmetric state}} \approx \hat{\psi}_{12,4}^{\text{vacuum}} \;, \quad \text{where} \quad \hat{\psi}_{12,4} = \left| \psi_{12,4}(x = 0) \right| \;.
\end{equation}  
Note also that due to the strong convective flux, the flattening of the pressure profile on the resonant surface is significantly more pronounced in the rotating state compared with the symmetric state (cf.\ Figs.\ \ref{fig:solution1}a and \ref{fig:solution2}a). Two dimensional maps of the pressure $p$ and the non-axisymmetric part of potential $\phi - \bar{\phi}$ corresponding to the two different equilibrium states are shown in Fig.\ \ref{fig:poloidal_cuts}. The periodic structure that corresponds to the flattening of the pressure on the magnetic islands is symmetric in the symmetric state (Fig.\ \ref{fig:poloidal_cuts}a) and distorted in the rotation state (Fig.\ \ref{fig:poloidal_cuts}b) from the shear flow. The potential structure together with the pressure structure, as illustrated in Fig.\ \ref{fig:poloidal_cuts}d, is responsible  for the strong convective flux in the rotation state. Note that with increasing amplitude of the external RMP current, both the magnetic flutter flux and the convective flux increase in the stable rotating equilibrium, but the convective flux is always larger than the thermal flux produced by magnetic flutters as shown in Fig.\ \ref{fig:compare_fluxes}. 
%
\section{Transition to the rotating state with strong convective flux}
\label{sec:rotation_transition}

\begin{figure}
\centerline{\includegraphics[width=0.5\textwidth]{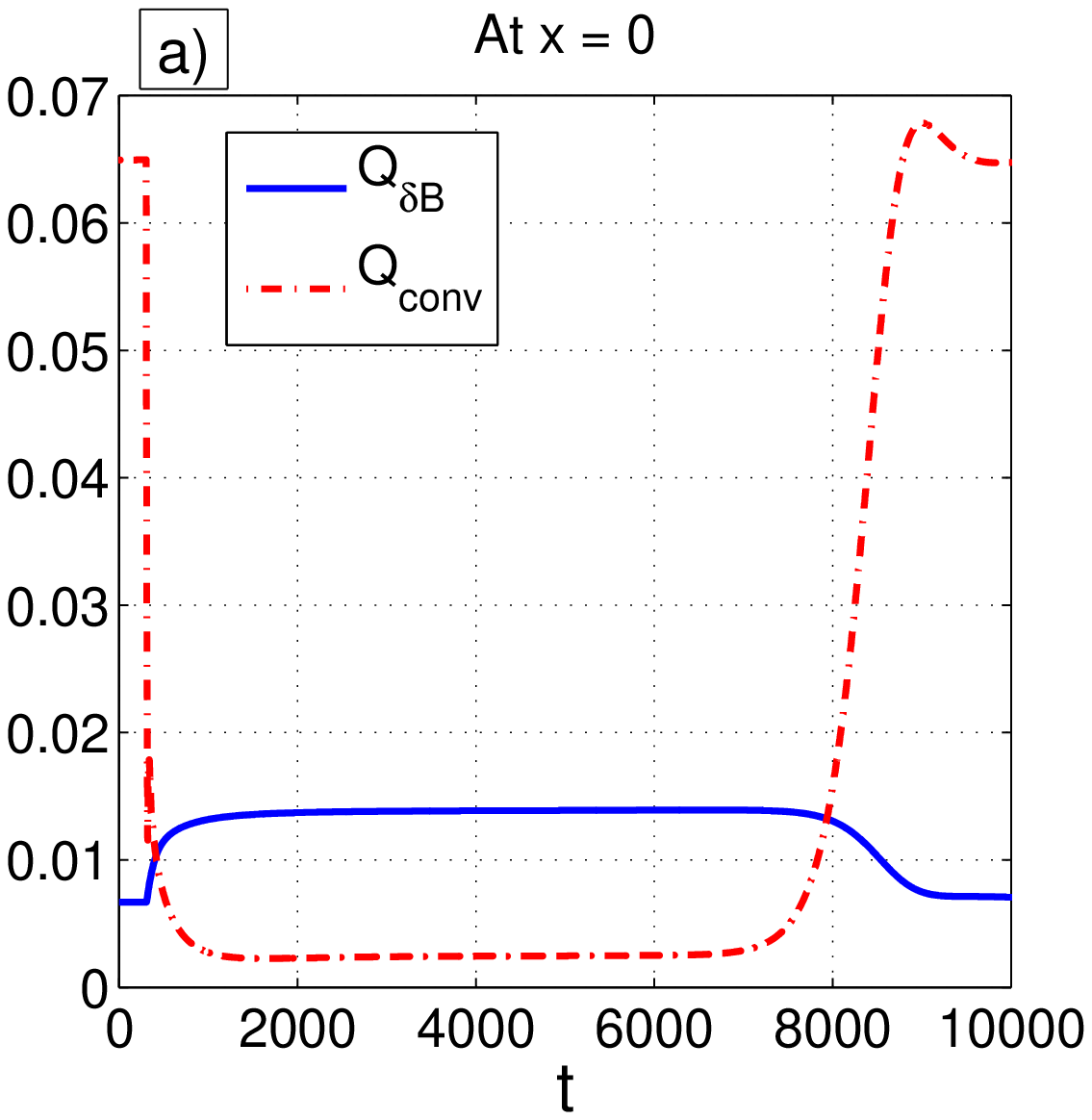}
\hfill\includegraphics[width=0.5\textwidth]{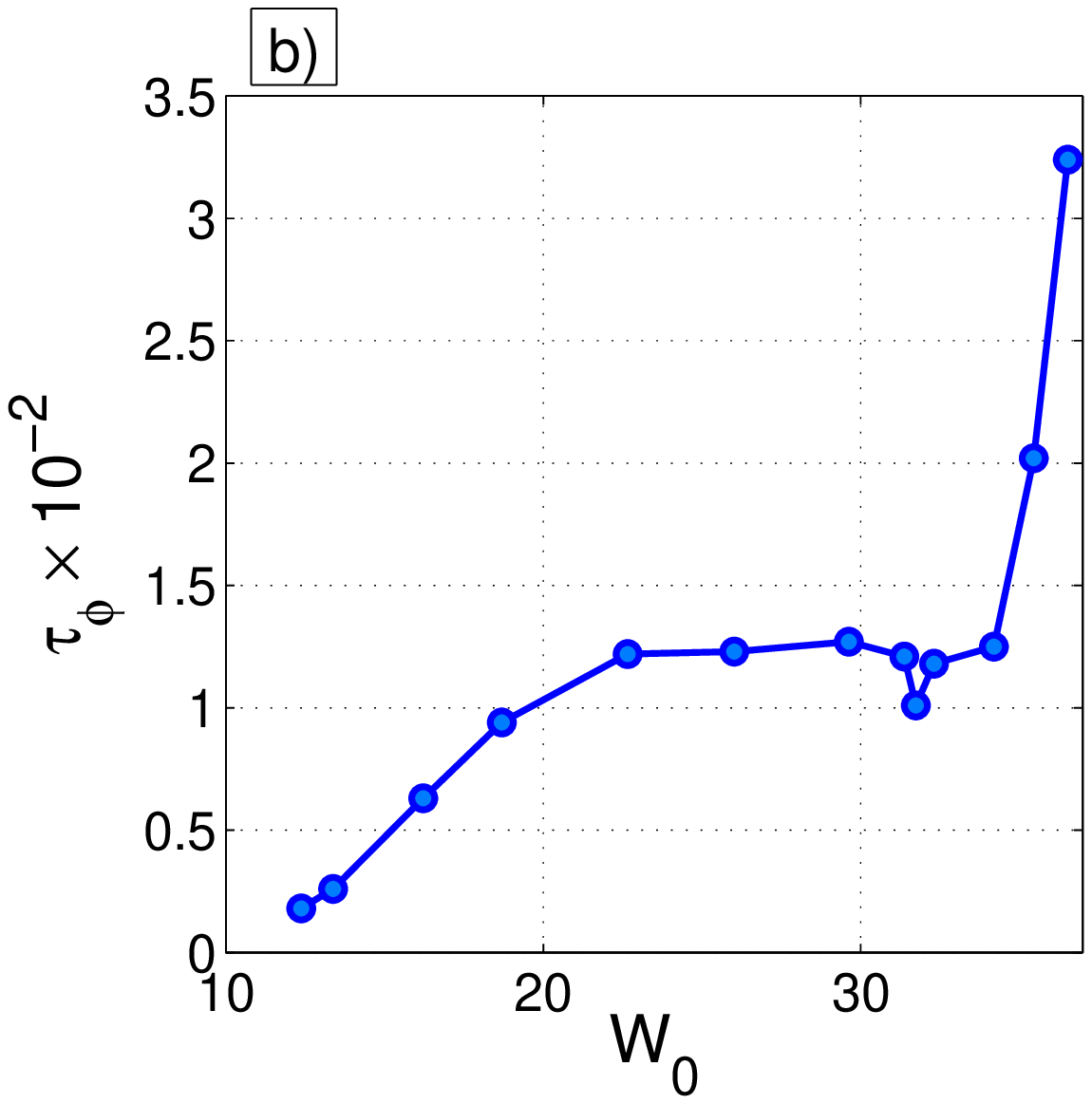}}
\caption{(Color online) - (a) Evolution of the convective (red dot dashed line) and magnetic flutter (blue line) fluxes. Forcing the rotation to zero, the convective flux decreases and the system responds increasing the flux of the magnetic flutter. Releasing the constrain, the system evolves to a stable condition with $Q_{conv}$ bigger than $Q_{\delta B}$. (b) The growth rate of the convective flux plotted against the magnetic island width $W_0$. \label{fig:flux_evolution}}
\end{figure}

\begin{figure}
\centerline{\includegraphics[width=0.5\textwidth]{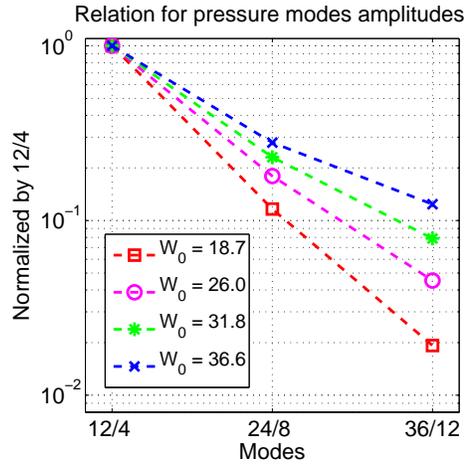}}
\caption{(Color online) Relation for resonant pressure modes amplitudes. Larger the island width $W_0$, which means higher resonant magnetic perturbation, more intense the harmonics modes become, contributing to the increase of $Q_{conv}$ and poloidal rotation effects.    \label{fig:higher_order}}
\end{figure}

The instability of the symmetric equilibrium and the transition to the second
helical state described above can be illustrated by performing a time
integration in two successive phases and following the evolution of the
convective and magnetic flutter fluxes at the resonant surface (Fig.\
\ref{fig:flux_evolution}a). In an early phase of the integration (from $t = 0.05
\cdot 10^4$ on), the rotation is forced to zero and the system is rapidly
evolving to the stationary state obtained in Eq.\ (\ref{eq:equilibrium}), where the convective flux $Q_\mathrm{conv}$ is much smaller than the flutter flux $Q_{\delta B}$.
Then, from $t = 8000$ on, we release the constraint on the poloidal
rotation and the system evolves self-consistently to the rotating state characterized by $Q_\mathrm{conv} \gg Q_{\delta B}$ .
The instability of the symmetric equilibrium can be characterized by the growth rate of a small perturbations of this equilibrium. The growth rate can be determined in the numerical experiment described above. A series of such numerical simulations for different amplitudes of the external magnetic perturbation reveals that the growth rate is nearly constant up to a critical value of the RMP perturbation level. Above this level, the growth rate is strongly increasing with the external perturbation amplitude. This is illustrated in Fig.\ \ref{fig:flux_evolution}b where the growth rate is plotted against the magnetic island width $W$ linked to the perturbation amplitude via
\begin{equation}
W = 4 \sqrt{2 \hat{\psi}_{12,4}} \;.
\end{equation}
The time growth rate of $Q_{conv}$ strongly increases for island widths $W > 34$. In that case, the half-width of the island $W / 2 > 17$ approaches the distance between the resonant surface and the external boundary of the main computational domain $x_{q=3.5} - x_{q=3} = 23$ and the island likely is influenced by the boundary. Also, higher order harmonics become significant for $W > W_c \approx 22$, where the critical island with is given by \cite{Fitz_PoP95} $W_c = \left[(8 / m_0)(r_0 / \xi)\right]^{1/2} \left( \chi_\perp / \chi_\parallel \right)^{1/4}$. As illustrated in Fig.\ \ref{fig:higher_order}, for $W = 18$ the amplitude of the second order $(m, n) = (24, 8)$ mode is one order of magnitude lower than the amplitude of the main harmonics but for $W = 36.5$, the second harmonic is only lower by a factor of $0.3$.
%
\section{Stable rotating state in toroidal geometry with multiple RMPs}
\label{sec:toroidal}

\begin{figure}
\centerline{\includegraphics[width=.45\textwidth]{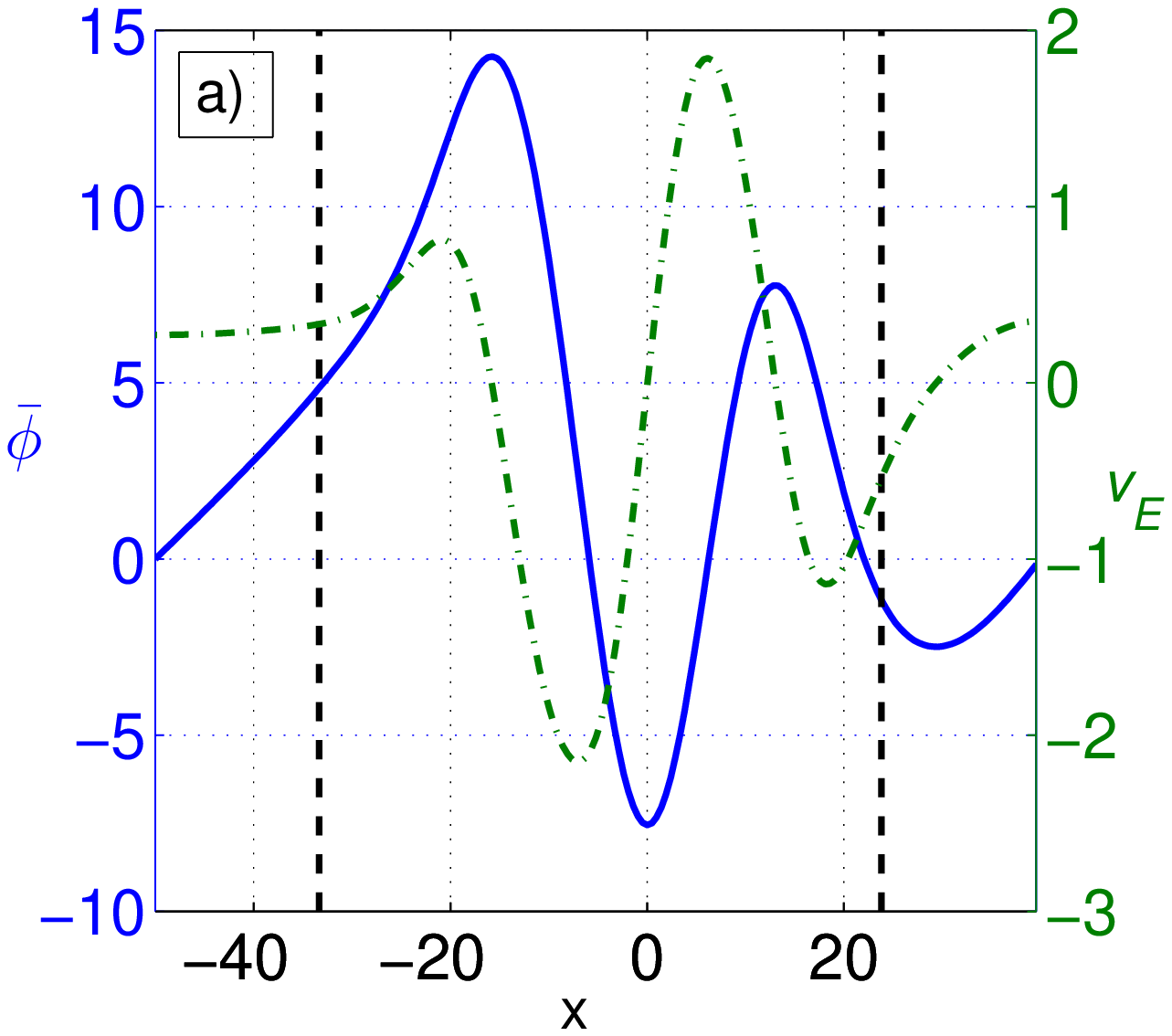}
\hfill\includegraphics[width=.45\textwidth]{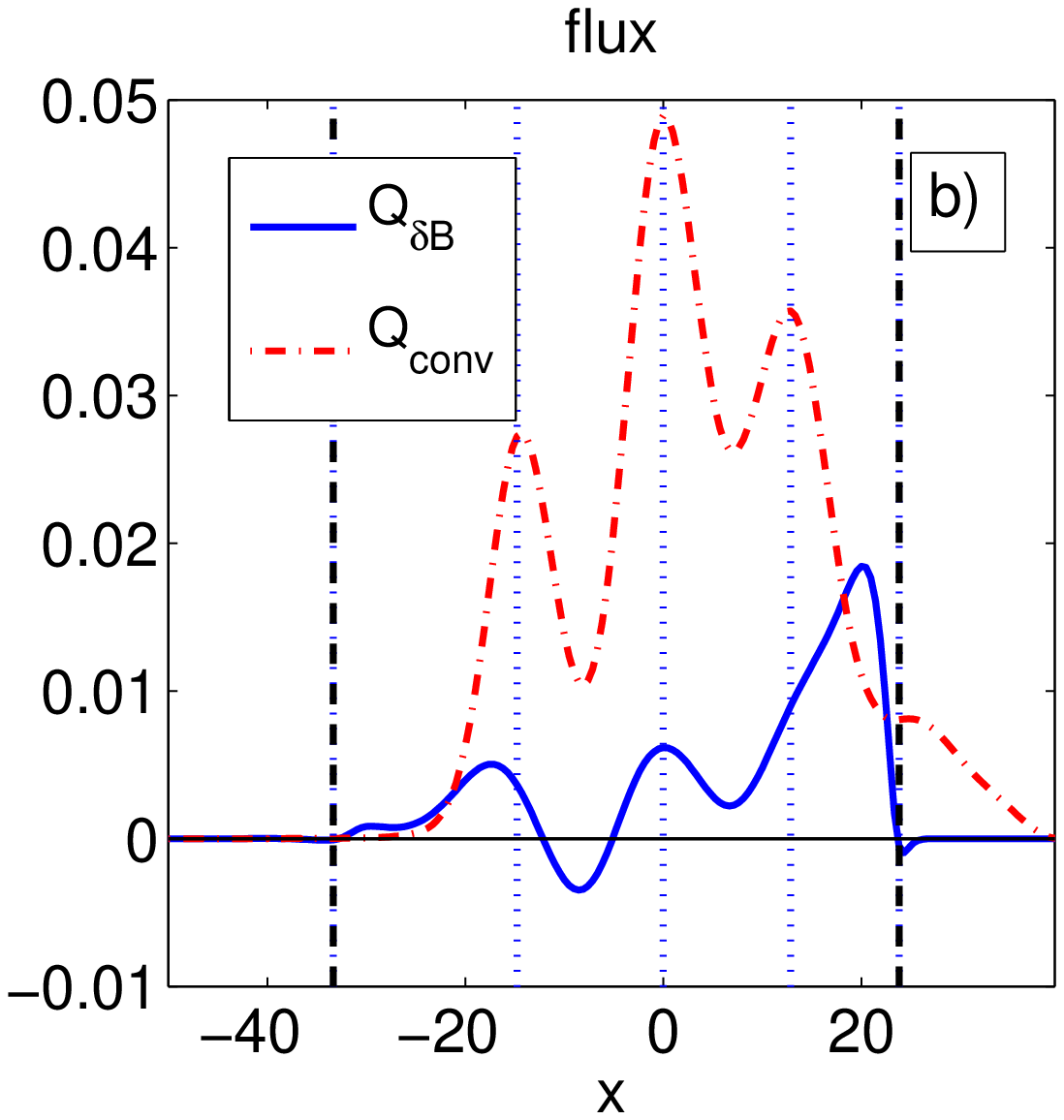}}
\caption{(Color online) Multiple harmonics RMP for cylindrical curvature with island width $W_0 = 18.7$. Picture a) depicts the equilibrium electric potential and poloidal rotation $v_E(x)$, and we find in b) the difference in the flux intensities for $Q_{conv}$ and $Q_{\delta B}$. \label{fig:multiple_cyl}}
\end{figure}

\begin{figure}
\centerline{\includegraphics[width=0.45\textwidth]{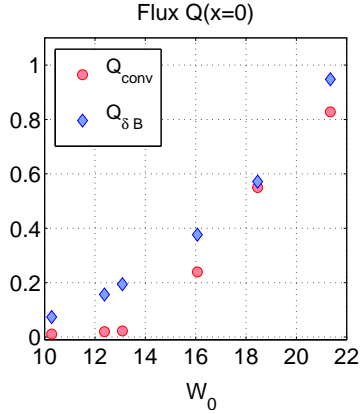}}
\caption{(Color online) Radially integrated convective and flutter fluxes for the toroidal case, $Q_\mathrm{conv} = \int_{q=2.5}^{q=3.5} Q_\mathrm{conv}(x)\,\mathrm{d}x$ and $Q_{\delta B} = \int_{q=2.5}^{q=3.5} Q_{\delta B}(x)\,\mathrm{d}x$, as a function of  multiple magnetic perturbations amplitudes expressed in terms of the vacuum island width $W_0$. For low values of external resonant magnetic perturbation $\psi_0$, which correspond to island width $W_0 < 13$ and Chirikov overlapping parameter $\sigma < 1$, $Q_{conv}$ is smaller than $Q_{\delta B}$ and the poloidal rotation is absent. \label{fig:compare_fluxes_toroidal}}
\end{figure}

\begin{figure}
\centerline{\includegraphics[width=.45\textwidth]{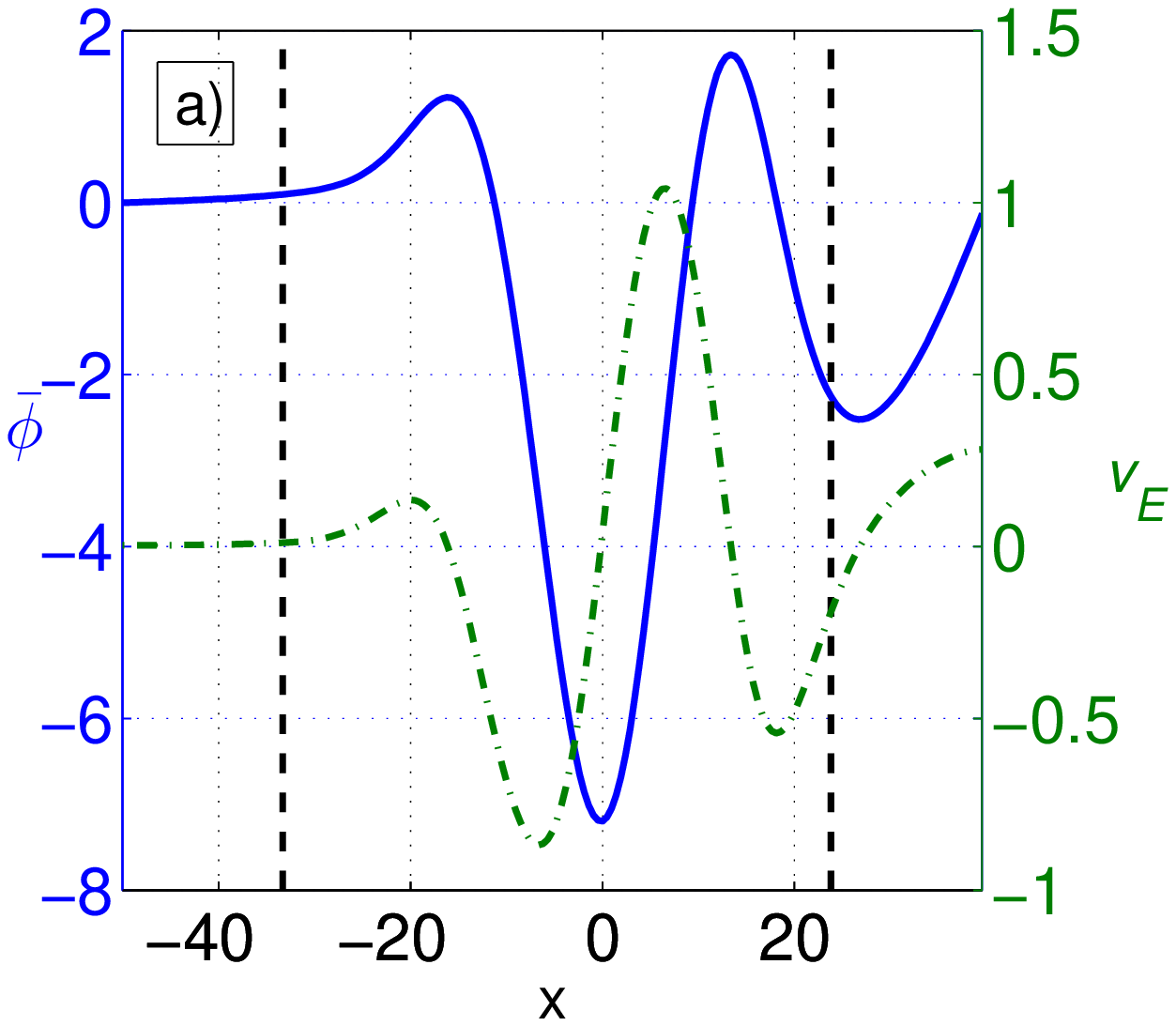}
\hfill\includegraphics[width=.45\textwidth]{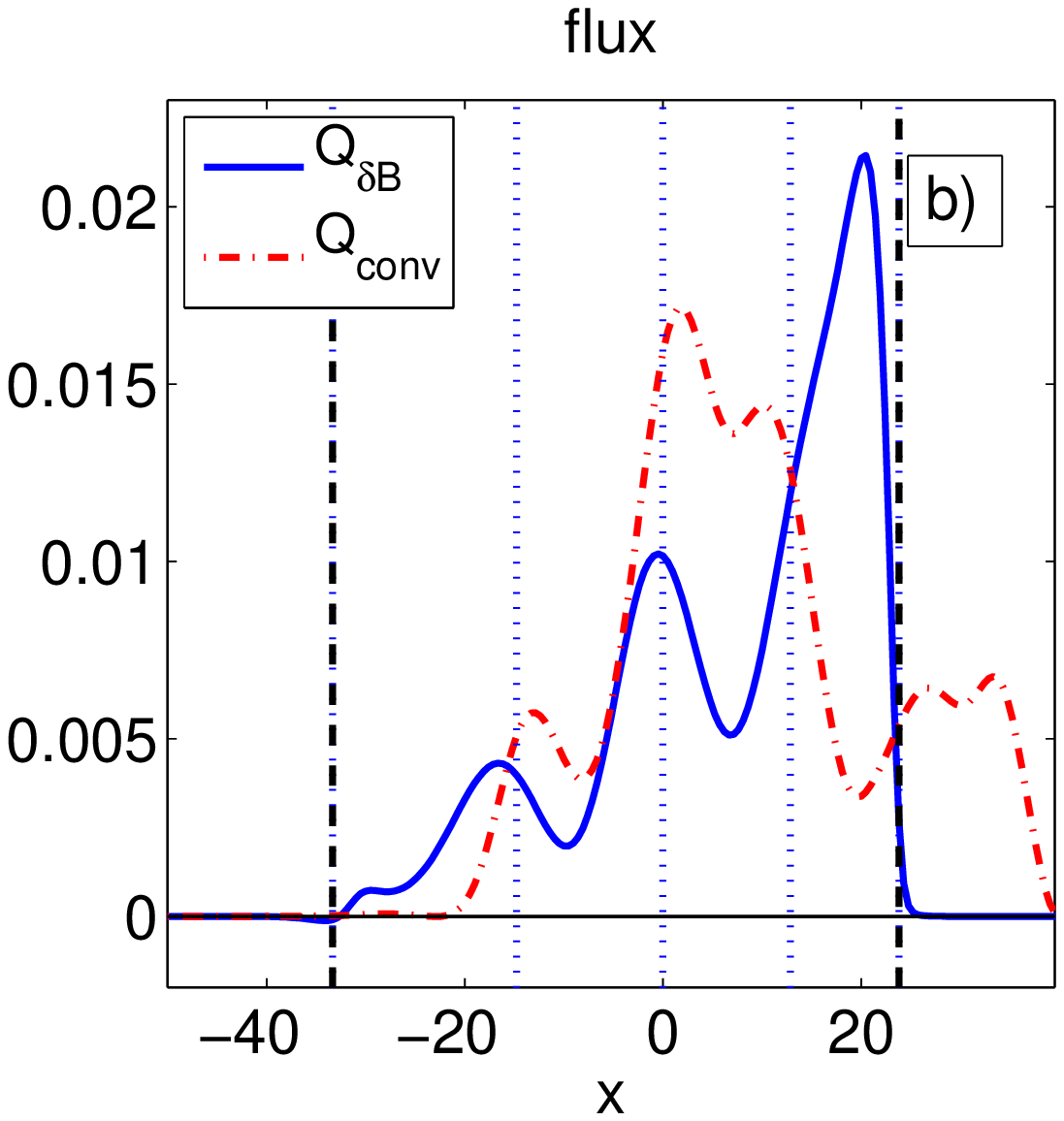}}
\caption{(Color online) Multiple harmonics RMP for toroidal curvature island width $W_0 = 18.7$. As described in Fig.\ \ref{fig:multiple_cyl}, a) depicts the equilibrium electric potential and poloidal rotation $v_E(x)$ and b) the difference in the flux intensities for $Q_{conv}$ and $Q_{\delta B}$.  \label{fig:multiple_rot}}
\end{figure}

Toroidal curvature gives rise to linear coupling between $m-1$, $m$, and $m+1$ modes. In particular, in the vorticity equation (\ref{eq:model_phi}) a $(m,n)$ harmonic of the pressure $p$ couples to $(m-1,n)$ and $(m+1,n)$ harmonics of the electrostatic potential $\phi$. If a self-consistent state with rotation and convective transport similar to the one shown in the Fig.\ \ref{fig:solution2} exists in toroidal geometry, then the state involve multiple harmonics. We therefore induce a multiple harmonics RMP with poloidal wavenumbers $m=10, 11, 12, 13, 14$ and toroidal wavenumber $n=4$. In order to compare with the results shown above, we first apply the multiple harmonics RMP in the cylindrical curvature case. As illustrated in Fig.\ \ref{fig:multiple_cyl}, a stable equilibrium with rotation and important convective flux is recovered in this case. The plasma self-organizes such that the rotation velocity vanishes close to the resonant surfaces $q= 11/4, 12/4, 13/4$. The convective and magnetic flutter fluxes show local maxima close to these resonant surfaces (Fig.\ \ref{fig:multiple_cyl}b). The width of the $(12,4)$ island is $W_{12,4} = 18.5$ and the width of the closest neighbouring island $(13,4)$ is $W_{13,4} = 18.4$, so the corresponding Chirikov overlapping parameter \cite{Chirikov} is
\begin{displaymath}
\sigma = \frac{W_{12,4} + W_{13,4}}{2 x_{q=3.25}} \approx 1.4 \;.
\end{displaymath}
The islands therefore are overlapping, but this overlapping is weak enough such that  distinct local maxima are observed for the $Q_{conv}$ and $Q_{\delta B}$ fluxes, as shown in Fig.\ \ref{fig:multiple_cyl}b.
When applying a multiple harmonics RMPs in the toroidal curvature case, the plasma also evolves to a stable equilibrium with rotation and strong convective flux. This is illustrated in Fig.\ \ref{fig:multiple_rot}, which may be compared with Fig.\ \ref{fig:multiple_cyl}. Differently from the cylindrical case in which a poloidal rotation is induced for all values of RMPs amplitudes, the poloidal rotation in toroidal geometry is triggered only for multiple harmonics RMPs and when Chirikov overlapping parameter $\sigma > 1$, which correspond to values of external resonant magnetic perturbation $\psi_0$ with island width $W_0 > 13$. In Fig.\ \ref{fig:compare_fluxes_toroidal} we see for the three first values of $W_0$, $Q_{conv}$ stays practically constant as the value of $Q_{\delta B}$ increases. This corresponds to the situation where the Chirikov overlapping parameter is smaller than one. 
Comparing figure \ref{fig:solution2},\ \ref{fig:multiple_cyl} and\  \ref{fig:multiple_rot} we see that even for different shapes and intensities of the equilibrium electric potential and convective fluxes, the poloidal rotation in all the three figures are at the same order of magnitude and have the maximum shear $\vert dv_E/dx \vert$ at the $q=3$ resonant surface. This indicates that an important induced poloidal rotation can be controlled by the RMP currents.    

\section{Conclusions}
\label{sec:conclusion}

In this work we investigate the dependence of the radial transport of the thermal energy from convection and magnetic flutter from the plasma response to a reference model of the resonant magnetic perturbations RMPs.  The stationary states are studied running the 3D plasma edge turbulence code EMEDGE3D below to primary ballooning instability threshold.
The simple static equilibrium in which the magnetic perturbation inside the plasma is in phase with the external perturbation and the plasma is not rotating is found to be unstable. The plasma is self-organized into a more complex state where the perturbation becomes phase shifted and the plasma rotates. This is due to the coupling between pressure and electrostatic potential perturbations induced by the magnetic curvature. In the stable equilibrium state, the phase between pressure and electrostatic potential is such that produces a rotating plasma with a significant thermal convective that exceeds the thermal flux from the magnetic flutter by a factor of 3--10 times.
The instability of the simple equilibrium and the subsequent evolution to a new stable complex equilibrium has been first investigated in cylindrical geometry for a single harmonic RMP perturbation and latter for multiple RMP modes. We showed that for both situations the system reaches the same final equilibrium state with high convective flux and induced poloidal rotation. Then, we showed that the corresponding coupling mechanism between pressure and the electric potential also produces plasma rotation and a strong convective flux in the toroidal geometry with multiple RMPs. The convective thermal flux and the magnetic flutter flux are about the same order of magnitude, but the presence of the thermal convective flux condition is only achieved when Chirikov overlapping parameter is greater than one.  

\begin{acknowledgments}
The authors gratefully acknowledge Wendell Horton and Iber\^e Caldas for useful discussions.
This research was supported by the French National Research Agency, project ANR-
2010-BLAN-940-01, and Brazilian national research agencies CNPq/CAPES under project No. 163395/2013-6 and FAPESP No. 201119296-1. 
\end{acknowledgments}  
%
%
\end{document}